\documentclass[onecollarge,final]{svjour2}          
\usepackage{graphicx}
\usepackage{color}
\usepackage{amsmath,amssymb}
\usepackage{mathptmx}      
\begin{document}

\title{Light Deflection and Gauss--Bonnet Theorem:
Definition of Total Deflection Angle and its Applications}

\author{Hideyoshi ARAKIDA}

\institute{
Hideyoshi. ARAKIDA \at
College of Engineering, Nihon University, Koriyama, Fukushima 963-8642 JAPAN\\
\email{arakida.hideyoshi@nihon-u.ac.jp}           
}


\maketitle

\begin{abstract}
In this paper, we re-examine the light deflection in the Schwarzschild 
and the Schwarzschild--de Sitter spacetime. First, supposing a static 
and spherically symmetric spacetime, we propose the definition of 
the total deflection angle $\alpha$ of the light ray by constructing 
a quadrilateral $\Sigma^4$ on the optical reference geometry 
${\cal M}^{\rm opt}$ determined by the optical metric $\bar{g}_{ij}$. 
On the basis of the definition of the total deflection angle $\alpha$ 
and the Gauss--Bonnet theorem, we derive two formulas to calculate 
the total deflection angle $\alpha$; (i) the angular formula that uses 
four angles determined on the optical reference geometry ${\cal M}^{\rm opt}$ 
or the curved $(r, \phi)$ subspace ${\cal M}^{\rm sub}$ being a slice 
of constant time $t$ and (ii) the integral formula on the optical reference 
geometry ${\cal M}^{\rm opt}$ which is the areal integral of 
the Gaussian curvature $K$ in the area of a quadrilateral $\Sigma^4$ and 
the line integral of the geodesic curvature $\kappa_g$ along 
the curve $C_{\Gamma}$. As the curve $C_{\Gamma}$, we introduce the 
unperturbed reference line that is the null geodesic $\Gamma$ on 
the background spacetime such as the Minkowski or the de Sitter spacetime, 
and is obtained by projecting $\Gamma$ vertically onto the curved 
$(r, \phi)$ subspace ${\cal M}^{\rm sub}$.
We demonstrate that the two formulas give the same total deflection
angle $\alpha$ for the Schwarzschild and the Schwarzschild--de Sitter
spacetime. In particular, in the Schwarzschild case, the result
coincides with Epstein--Shapiro's formula when the source $S$ and
the receiver $R$ of the light ray are located at infinity. 
In addition, in the Schwarzschild--de Sitter case, there appear order
${\cal O}(\Lambda m)$ terms in addition to the Schwarzschild-like
part, while order ${\cal O}(\Lambda)$ terms disappear.
\end{abstract}

keywords{gravitation \and cosmological constant \and light deflection}
%
%
%
%
%

\section{Introduction\label{sec:intro}}
In 1919, light deflection by the mass of the sun was detected
during the total solar eclipse in Sobral and Principe
\cite{dyson_etal1920}. These observations provided the
first evidence of the validity of the general theory of relativity,
as well as the stunning explanation of the perihelion advance of
Mercury. Thereafter, measurement of the bending of light played 
an important role in the verification of the general theory of relativity 
\cite{will2014}. Further, gravitational lensing which is based on light 
deflection is widely used as a powerful tool to investigate various 
astronomical/astrophysical phenomena such as dark matter exploration 
and extrasolar planet search, see e.g., 
\cite{schneider_etal1999,schneider_etal2006} and the references therein.

Because the method for calculating the total deflection angle
$\alpha$ is currently described in many textbooks and literature,
one may think that computing the total deflection angle is now an 
elementary problem and essential issues concerning the deflection of 
light have already been understood and solved completely.
However, there still exist some basic issues to be settled
that were highlighted when considering the contribution of
cosmological constant $\Lambda$ to the bending of light.

Islam \cite{islam1983} showed that the trajectory of the light ray
is not related to the cosmological constant $\Lambda$ because
the second-order differential equation of the photon (a null geodesic)
does not contain $\Lambda$. Therefore, it was considered
for a long time that $\Lambda$ does not contribute to the light
deflection. However, in 2007, Rindler and Ishak \cite{rindler2007}
pointed out that $\Lambda$ affects the bending angle of the light ray
by using the invariant cosine formula under the Schwarzschild--de
Sitter/Kottler solution. Starting with this paper \cite{rindler2007},
many authors intensively discussed its appearance in diverse ways;
see \cite{ishak2010} for a review article, and also see e.g.,
\cite{lake2002,park2008,kp2008,sph2010,bhadra2010,miraghaei2010,biressa2011,ak2012,hammad2013,lebedevlake2013,batic_etal2015,arakida2016} and the references therein.
However, despite various discussions and approaches, a definitive conclusion 
has not yet emerged. The main reasons are as follows:
\begin{enumerate}
 \item If there exists a cosmological constant $\Lambda$ such as the 
       Schwarzschild--de Sitter solution, the spacetime is not asymptotically 
       flat unlike the Schwarzschild spacetime, and we cannot then apply 
       the standard procedure described in many textbooks and literature.
 \item In the first place, it is ambiguous and not clear
       what is the total deflection angle and how it should be defined.
\end{enumerate}

The application of the Gauss--Bonnet theorem has the potential to solve 
above problems and settle the arguments by examining the total deflection 
angle correctly in the curved spacetime such as the Schwarzschild--de 
Sitter/Kottler solution. The Gauss--Bonnet theorem has already been 
adopted in previous papers e.g., 
\cite{gibbons_werner2008a,gibbons_werner2008b,ishihara_etal2016,ishihara_etal2017}.
Although these pioneering works, especially \cite{ishihara_etal2016}, 
have been successful in showing effectiveness of the Gauss--Bonnet theorem, 
it seems that they have not reached the stage of giving a definite definition 
of the total deflection angle and some points remain to be clarified.

Not only from a theoretical viewpoint but also from the standpoint of
an observational evidence, the cosmological constant problem is important 
(See reviews by e.g., \cite{weinberg1989,carroll2001}).
Recent cosmological observations strongly suggest the acceleration 
of cosmic expansion. Today, it is widely considered that the cosmological
constant $\Lambda$, or the dark energy in a more general sense,
is the most promising candidate to explain the observed accelerating 
expansion of the universe although the details are completely unknown 
so far. As one way to tackle this problem from another perspective, 
it is natural to investigate the effect of the cosmological constant 
$\Lambda$ on the classical tests of the general theory of relativity, 
especially on the deflection of the light ray.

In this paper, first we will propose a definition for total deflection 
angle $\alpha$ of the light ray presupposing the static and spherically 
symmetric spacetime. After introducing a definition for total deflection 
angle we will develop two methods for calculating the total deflection 
angle on the basis of the Gauss--Bonnet theorem. These two formulas will 
be applied to the Schwarzschild and the Schwarzschild--de Sitter spacetime, 
and we will demonstrate that they give equivalent results. In particular, 
in the Schwarzschild case, the result coincides with Epstein--Shapiro's
formula when the source $S$ and the receiver $R$ of the light ray are 
located at infinity. In addition, in the case of the Schwarzschild--de 
Sitter spacetime, our results indicate that the cosmological constant 
$\Lambda$ contributes to the total deflection angle as the form of 
the ${\cal O}(\Lambda m)$ terms (here $m$ is the mass of the central body), 
while the ${\cal O}(\Lambda)$ terms do not appear; this is connected with 
the fact that the light ray does not bend in the de Sitter spacetime.

This paper is organized as follows; first, the optical metric is
introduced in section \ref{sec:optical} and the outline of
the Gauss--Bonnet theorem is presented in section \ref{sec:GBtheorem}.
The total deflection angle is defined, and on the basis of the
definition of the total deflection angle 
we construct two formulas for calculating the total deflection
angle in section \ref{sec:defangle}.
The two formulas are applied to an asymptotically flat spacetime
in section \ref{sec:asymptotic} and to a non-asymptotically flat
spacetime in section \ref{sec:non-asymptotic}. Finally, section
\ref{sec:conclusion} concludes the paper.

\section{Optical Metric\label{sec:optical}}
We assume that the spacetime is expressed in the static and
spherically symmetric form
\begin{eqnarray}
 ds^2 = g_{\mu\nu}dx^{\mu}dx^{\nu}
  =  -f(r)dt^2 + \frac{1}{f(r)}dr^2
  + r^2(d\theta^2 + \sin^2\theta d\phi^2),
  \label{eq:metric1}
\end{eqnarray}
where $f(r)$ is a function of the radial coordinate $r$,
a Greek subscript such as $\mu, \nu$ runs from 0 to 3, 
and we choose the geometrical unit $c = G = 1$.
Because of the spherical symmetry, without losing generality,
it is possible to take the equatorial plane $\theta = \pi/2, d\theta = 0$
as the orbital plane of the light rays
\begin{eqnarray}
 ds^2 = -f(r)dt^2 + \frac{1}{f(r)}dr^2 + r^2 d\phi^2.
  \label{eq:metric2}
\end{eqnarray}
We have two constants of motion, that is, the energy $E$ and
the angular momentum $L$,
\begin{eqnarray}
 E = f(r)\frac{dt}{d\lambda},\quad
  L = r^2\frac{d\phi}{d\lambda},
  \label{eq:constant1}
\end{eqnarray}
in which $\lambda$ is an affine parameter. From two constants of motion 
$E$ and $L$, another constant, the so-called the impact parameter $b$, 
is determined by
\begin{eqnarray}
 b \equiv \frac{L}{E}.
  \label{eq:impact}
\end{eqnarray}
Further, we derive the following relation from Eqs. (\ref{eq:constant1})
and (\ref{eq:impact}),
\begin{eqnarray}
 \frac{d\phi}{dt} = \frac{bf(r)}{r^2}.
  \label{eq:dpdt}
\end{eqnarray}

The light ray satisfies the null condition $ds^2 = 0$, and
from this null condition let us introduce the optical metric
$\bar{g}_{ij}$ which can be regarded as the Riemannian geometry
experienced by the light rays: 
\begin{eqnarray}
 dt^2 \equiv \bar{g}_{ij} dx^idx^j
 = \bar{g}_{rr}dr^2 + \bar{g}_{\phi\phi}d\phi^2
 = \frac{1}{[f(r)]^2}dr^2 + \frac{r^2}{f(r)}d\phi^2,
  \label{eq:optical1}
\end{eqnarray}
in which a Latin subscript such as $i, j$ takes $i, j = 1, 2$ and
means $1 = r$ and $2 = \phi$. Hereafter we call the geometry
defined by the optical metric $\bar{g}_{ij}$ the optical reference
geometry ${\cal M}^{\rm opt}$ in accordance with \cite{abramowicz1988}.

On the other hand, taking the slice of constant time $t$ of 
the spacetime Eq. (\ref{eq:metric2}), a two-dimensional curved $(r, \phi)$ 
space (the spatial part of Eq. (\ref{eq:metric2})) is described by
\begin{eqnarray}
 d\ell^2 \equiv g_{ij}dx^idx^j
 =
  g_{rr}dr^2 + g_{\phi\phi}d\phi^2
 =
  \frac{dr^2}{f(r)} + r^2d\phi^2,
  \label{eq:metric3}
\end{eqnarray}
and we call this geometry determined by $g_{ij}$ the curved $(r, \phi)$ 
subspace ${\cal M}^{\rm sub}$.

We mention that Eqs. (\ref{eq:optical1}) and (\ref{eq:metric3}) are 
connected by the conformal transformation
\begin{eqnarray}
 \bar{g}_{ij} = \omega^2 (x)g_{ij},
  \label{eq:metric4}
\end{eqnarray}
or more generally,
\begin{eqnarray}
  \bar{g}_{\mu\nu} = \omega^2 (x) g_{\mu\nu},
\end{eqnarray}
where $\omega^2(x)$ is called the conformal factor which in our case is
\begin{eqnarray}
 \omega^2 (x) = \frac{1}{f(r)}.
\end{eqnarray}
It is noteworthy that the conformal transformation preserves the angle of 
the point at which the two curves intersect, and rescales the coordinate value. 
Furthermore the null geodesic does not change its form by the 
conformal transformation because of the null condition; 
see e.g., Appendix G in \cite{carroll2004}.

Setting the unit tangent vector $k^i$ of light ray path on
${\cal M}^{\rm opt}$ as
\begin{eqnarray}
 k^i = \frac{dx^i}{dt},
\end{eqnarray}
and from Eq. (\ref{eq:optical1}), $k^i$ is actually unit vector:
 \begin{eqnarray}
  1 = \bar{g}_{ij}k^i k^j.
   \label{eq:optical2}
 \end{eqnarray}
It should be noticed that on the optical reference geometry 
${\cal M}^{\rm opt}$, $t$ plays the role of an arc length parameter because
\begin{eqnarray}
 \int_{t_1}^{t_2} dt =
  \int_{t_1}^{t_2}
  \sqrt{\bar{g}_{rr}(k^r)^2 + \bar{g}_{\phi\phi}(k^{\phi})^2}dt
  = t_2 - t_1.
  \label{eq:optical3}
\end{eqnarray}
Therefore, the optical reference geometry ${\cal M}^{\rm opt}$ 
is suitable for application to the Gauss--Bonnet theorem later. 

Finally, in accordance with \cite{ishihara_etal2016}, we prepare
the radial and tangential unit vector $e_r^i$, $e_{\phi}^i$ as
\begin{eqnarray}
 e_r^i = \left(\frac{1}{\sqrt{\bar{g}_{rr}}}, 0\right),\quad
 e_{\phi}^i = \left(0, \frac{1}{\sqrt{\bar{g}_{\phi\phi}}}\right).
\end{eqnarray}
It is easy to check that these two vectors are certainly
the unit vectors because of
\begin{eqnarray}
\bar{g}_{ij}e_r^i e_r^j = 1,\quad
 \bar{g}_{ij} e_{\phi}^i e_{\phi}^j = 1.
\end{eqnarray}
Because we are working in the optical reference geometry ${\cal M}^{\rm opt}$, 
the inner product of the two vectors is defined by using the optical metric 
$\bar{g}_{ij}$ instead of $g_{ij}$.
\section{Gauss--Bonnet Theorem\label{sec:GBtheorem}}
We suppose that the line element is written in a diagonal form determined 
by the optical metric $\bar{g}_{ij}$; see Eq. (\ref{eq:optical1}).
Due to Eq. (\ref{eq:optical3}), the time $t$ plays the role of an 
arc length parameter in the optical reference geometry ${\cal M}^{\rm opt}$, 
and hereafter we denote an arc length parameter by $t$ instead of $s$ 
which is usually used.

\begin{figure}[htbp]
\begin{center}
 \includegraphics[scale=0.2,clip]{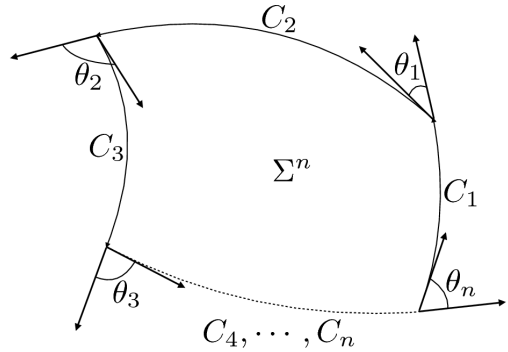}
 \caption{Schematic diagram of Gauss--Bonnet theorem.
 A polygon $\Sigma^n$ is bounded by the curves $C_1, C_2, \cdots, C_n$,
 and the external angles of a polygon $\Sigma^n$ are denoted by
 $\theta_p ~ (p = 1, 2, \cdots, n)$.\label{fig:arakida-fig1}}
\end{center}
\end{figure}

Consider a polygon $\Sigma^n$ with $n$ vertices on ${\cal M}^{\rm opt}$
which is orientable and bounded by $n$ smooth piecewise regular
curves $C_p ~ (p = 1, 2, \cdots, n)$; see Fig. \ref{fig:arakida-fig1}.
The (local) Gauss--Bonnet theorem is expressed as, e.g., on p. 139 
in \cite{klingenberg1978}, p. 170 in \cite{kreyszig1991}, and 
p. 272 in \cite{carmo2016} 
\begin{eqnarray}
 \iint_{\Sigma^n} K d\sigma
  +
 \sum_{p = 1}^n \int_{C_p} \kappa_g dt
  + \sum_{p = 1}^{n}\theta_p = 2\pi,
  \label{eq:GB1}
\end{eqnarray} 
in which an arc length parameter $t$ moves along the curve $C_p$ in such 
a sense that a polygon $\Sigma^n$ stays on the left side,
$d\sigma = \sqrt{\det|\bar{g}|}dx^1 dx^2 = \sqrt{\det|\bar{g}|}drd\phi$
is an areal element, and $\theta_p$ is the external angle at the 
$p$-th vertex which is described as the sense leaving the internal angle
on the left.
$K$ is the Gaussian curvature defined as, e.g., on p. 147
in \cite{kreyszig1991}
\begin{eqnarray}
 K = -\frac{1}{\sqrt{\bar{g}_{rr} \bar{g}_{\phi\phi}}}
  \left[
   \frac{
   \partial}{\partial r}
   \left(\frac{1}{\sqrt{\bar{g}_{rr}}}
    \frac{\partial\sqrt{\bar{g}_{\phi\phi}}}{\partial r}\right)
   +
   \frac{\partial}{\partial \phi}
   \left(\frac{1}{\sqrt{\bar{g}_{\phi\phi}}}
    \frac{\partial\sqrt{\bar{g}_{rr}}}{\partial \phi}\right)
	 \right].
  \label{eq:GB2}
\end{eqnarray}
The Gaussian curvature $K$ represents how surface is curved.
$\kappa_g$ is the geodesic curvature along the curve $C_p$,
e.g., on p. 256 in \cite{carmo2016}
\begin{eqnarray}
 \kappa_g = \frac{1}{2\sqrt{\bar{g}_{rr} \bar{g}_{\phi\phi}}}
  \left(
   \frac{\partial \bar{g}_{\phi\phi}}{\partial r}
   \frac{d\phi}{dt}
   -
   \frac{\partial \bar{g}_{rr}}{\partial \phi}
   \frac{dr}{dt}
	 \right)
  + \frac{d\Phi}{dt},
  \label{eq:GB3}
\end{eqnarray}
where $\Phi$ is the angle between the radial unit vector $e_r^i$
and the tangent vector of the curve $C_p$ in our case
\footnote{
In General, $\Phi$ is determined as an angle between
the tangent vector of the curve $C_p$ and the vector
$\partial S^i(u, v)/\partial u$, where $S^i$ means the point
on the surface $S$. In our case, $u = r, v = \phi$ then
$\partial S^i(r, \phi)/\partial r$ can be understood as the
vector pointing in the $r$ direction. The unit vector,
\[
 \frac{\frac{\partial S^i(r, \phi)}{\partial r}}
{\left\|\frac{\partial S^i(r, \phi)}{\partial r}\right\|},
\]
corresponds to $e_r^i$ in our notation.}.
The geodesic curvature $\kappa_g$ indicates how far the curve 
$C_p$ deviates from the geodesic. It should be mentioned if the 
curve $C_p$ is the geodesic, then $\kappa_g = 0$. 
We note that because we are working in the optical
reference geometry ${\cal M}^{\rm opt}$, $K$ and $\kappa_g$ are determined
in terms of the optical metric $\bar{g}_{ij}$ instead of $g_{ij}$.
\section{Definition of Total Deflection Angle\label{sec:defangle}}
\subsection{Unperturbed Reference Line\label{sec:rsl}}
In order to define the total deflection angle on the basis of
the Gauss--Bonnet theorem, we must prepare a polygon $\Sigma^n$ on 
the optical reference geometry ${\cal M}^{\rm opt}$.
The most natural way to do this is to construct a triangle $\Sigma^3$
with three points; the center $O$ where the massive star or galaxy
with mass $m$ is located, the source of light ray $S$ and the
receiver of the light ray $R$. However, in the case of the
Schwarzschild and the Schwarzschild--de Sitter spacetime,
the center $O$ is a singularity, so that it is not relevant
to constructing a triangle $\Sigma^3$ on ${\cal M}^{\rm opt}$.
Therefore, instead of a triangle $\Sigma^3$, let us configure
a quadrilateral $\Sigma^4$ on ${\cal M}^{\rm opt}$ that is bounded by 
the trajectory of the light rays, $\gamma$, two radial null geodesics
$\gamma_S$ connecting the center $O$ and the source of light ray $S$ and 
$\gamma_R$ connecting the center $O$ and the receiver of light ray $R$, 
and the curve $C$. So far, it is not clear what is the curve $C$; 
however if the curve $C$ is determined appropriately, then we can lay 
a quadrilateral $\Sigma^4$ on ${\cal M}^{\rm opt}$.

To seek the curve $C$, we start from the investigation of the curved
$(r, \phi)$ subspace ${\cal M}^{\rm sub}$ determined by 
Eq. (\ref{eq:metric3}) before considering the optical reference geometry 
${\cal M}^{\rm opt}$ determined by Eq. (\ref{eq:optical1}).
We recall the fact that if the curved $(r, \phi)$ subspace 
${\cal M}^{\rm sub}$ is restricted in the spherically symmetric form as
Eq. (\ref{eq:metric3}), two curved $(r_1, \phi_1)$ and
$(r_2, \phi_2)$ subspaces ${\cal M}^{\rm sub}_1$ and ${\cal M}^{\rm sub}_2$, 
characterized by $f_1(r_1)$ and $f_2(r_2)$, respectively, 
have the same coordinate values $r_1 = r_2$ and $\phi_1 = \phi_2$
because $r_1$ and $r_2$ are determined as the circumference radius
$\ell = 2\pi r_1 = 2\pi r_2$ for the constant $r$ (then $dr = 0$).
Further, the same relation also holds between the curved
$(r, \phi)$ subspace ${\cal M}^{\rm sub}$ and the flat $(\rho, \varphi)$ 
plane of the three-dimensional flat space with cylindrical coordinates
$(\rho, \varphi, z)$, $r = \rho$ and $\phi = \varphi$ because 
$\ell = 2\pi r = 2\pi \rho$ for constant $r$ ($dr = 0$) and $\rho$
($d\rho = 0$); see Appendix \ref{append:space} and also section 
11.3 in \cite{rindler2006} and section II in \cite{lebedevlake2013}.

\begin{figure}[htbp]
\begin{center}
 \includegraphics[scale=0.3]{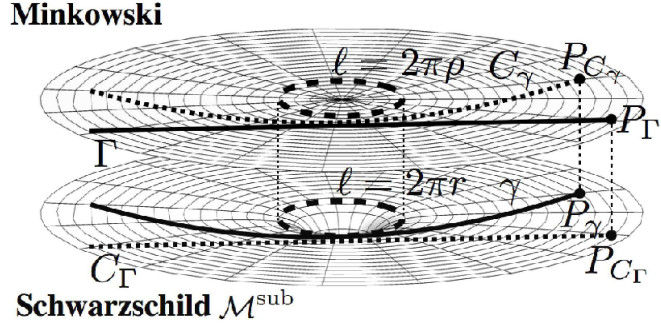}
 \caption{The relation between the Minkowski $(\rho, \varphi)$
 flat space and the Schwarzschild curved $(r, \phi)$ subspace
 ${\cal M}^{\rm sub}$.
 The null geodesics $\Gamma, \gamma$ are represented by solid lines, 
 the curves $C_{\Gamma}, C_{\gamma}$ are represented by dotted lines, and
 the two circumferences $\ell$ are represented by dashed lines.
 The points $P_{\Gamma}$ on $\Gamma$ and $P_{C_{\Gamma}}$ 
 on $C_{\Gamma}$ have the same coordinate values and 
 the points $P_{\gamma}$ on $\gamma$ and $P_{C_{\gamma}}$ 
 on $C_{\gamma}$ also have the same coordinate values
 because the circumference radius $\ell = 2\pi r = 2\pi \rho$
 and the vertical projection.
 \label{fig:arakida-fig2}}
\end{center}
\end{figure}

First, we examine the relation between the Schwarzschild and
the Minkowski spacetime. The Minkowski spacetime can be considered
as the background spacetime of the Schwarzschild one.
Because the Minkowski space coincides with the flat
$(\rho, \varphi)$ plane, we describe the Minkowski space
by using $(\rho, \varphi)$ coordinates.
Fig. \ref{fig:arakida-fig2} shows the relation between
the Minkowski $(\rho, \varphi)$ flat space and the Schwarzschild curved
$(r, \phi)$ subspace ${\cal M}^{\rm sub}$.
The differential equation of light ray $\Gamma$ in the
Minkowski $(\rho, \varphi)$ flat space is
\begin{eqnarray}
 \left.\left(\frac{d\rho}{d\varphi}\right)^2\right|_{\Gamma}
  = \rho^2_{\Gamma}
  \left(
   \frac{\rho^2_{\Gamma}}{b^2} - 1
   \right),
\end{eqnarray}
and it yields the equation of the trajectory of the light ray:
\begin{eqnarray}
 \frac{1}{\rho_{\Gamma}} = \frac{\sin\varphi}{b},
  \label{eq:Gamma1}
\end{eqnarray}
where $b$ is the impact parameter defined by Eq. (\ref{eq:impact}).
The form of the trajectory of the light ray, Eq. (\ref{eq:Gamma1}), is
a straight line in the Euclidean sense. Whereas, in the Schwarzschild case, 
the differential equation of light ray $\gamma$ on the curved 
$(r, \phi)$ subspace ${\cal M}^{\rm sub}$ is
\begin{eqnarray}
 \left.\left(\frac{dr}{d\phi}\right)^2\right|_{\gamma}
  = r^2_{\gamma}
  \left(\frac{r^2_{\gamma}}{b^2} + \frac{2m}{r_{\gamma}} - 1\right),
\end{eqnarray}
and $r_{\gamma}$ can be expressed by means of $\phi$ as the form
\begin{eqnarray}
 r_{\gamma} = r_{\gamma}(\phi; m, b).
  \label{eq:gamma1}
\end{eqnarray}
For the concrete form of $r_{\gamma}$, see e.g., Eq. (\ref{eq:Schtrajectory1}).
Eq. (\ref{eq:gamma1}) is a null geodesic $\gamma$ on the curved $(r, \phi)$
subspace ${\cal M}^{\rm sub}$. However, because of the relation
\begin{eqnarray}
 r = \rho,\quad \phi = \varphi,
\end{eqnarray}
the form of Eq. (\ref{eq:gamma1}) does not change on the Minkowski 
$(\rho, \varphi)$ flat space; however it is not geodesic $\gamma$ 
but just a curve $C_{\gamma}$ which is vertical projection of $\gamma$ 
onto the Minkowski $(\rho, \varphi)$ flat space.

We mention that the two geodesics $\Gamma$ on the Minkowski $(\rho, \varphi)$ 
flat space and $\gamma$ on the curved $(r, \phi)$ subspace ${\cal M}^{\rm sub}$ 
(the Schwarzschild space) are the straight line in such a sense 
that according to Fermat's principle, they take the shortest path 
between two points in the shortest time. Therefore, both $\Gamma$ on 
the Minkowski $(\rho, \varphi)$ flat space and $\gamma$ on 
the curved $(r, \phi)$ subspace ${\cal M}^{\rm sub}$ (the Schwarzschild space)
do {\it not} bend; in fact, the line integral of geodesic curvature $\kappa_g$
along the null geodesics $\Gamma$ and $\gamma$ vanish (please remind the
meaning of the geodesic and the geodesic curvature $\kappa_g$).

In order to obtain the total deflection angle $\alpha$, we must compare 
the null geodesics $\Gamma$ and $\gamma$ existing in the distinct spaces. 
For this purpose, we usually project $\gamma$ vertically onto 
the Minkowski $(\rho, \varphi)$ flat space and regard the difference 
in the direction of light trajectory between the null geodesic $\Gamma$ 
and the curve $C_{\gamma}$ as the total deflection angle $\alpha$; 
see Fig. \ref{fig:arakida-fig2}. However, this approach works only when 
the source $S$ and the receiver $R$ of light ray are located at infinity
because the Schwarzschild curved $(r, \phi)$ subspace becomes asymptotically flat,
corresponding to the Minkowski $(\rho, \varphi)$ flat space.
Hence, if the source $S$ and/or the receiver $R$ of light ray are
placed within a finite region on the curved $(r, \phi)$ subspace
${\cal M}^{\rm sub}$, we cannot determine the total deflection angle
on the Minkowski $(\rho, \varphi)$ flat space anymore.

To overcome this situation, let us try to determine the total deflection angle 
$\alpha$ on the Schwarzschild curved $(r, \phi)$ subspace ${\cal M}^{\rm sub}$
because the actual light trajectory lies on the curved $(r, \phi)$ 
subspace ${\cal M}^{\rm sub}$. Due to the relation $r = \rho$ and $\phi = \varphi$,
Eq. (\ref{eq:Gamma1}) holds in the same form on the Schwarzschild curved 
$(r, \phi)$ subspace ${\cal M}^{\rm sub}$. In this case, Eq. (\ref{eq:Gamma1}) 
is no longer the geodesic $\Gamma$ and is just a curve $C_{\Gamma}$ which is 
the vertical projection of $\Gamma$ onto the Schwarzschild curved $(r, \phi)$ subspace:
\begin{eqnarray}
 \frac{1}{r_{C_{\Gamma}}} = \frac{\sin\phi}{b}.
  \label{eq:Gamma2}
\end{eqnarray}
It can be said that to evaluate $C_{\Gamma}$ on the curved $(r, \phi)$
subspace ${\cal M}^{\rm sub}$ means to measure $C_{\Gamma}$ 
with a curved metric (ruler) instead of with a flat metric (ruler).
By using Eqs. (\ref{eq:gamma1}) and (\ref{eq:Gamma2}),
we can construct a quadrilateral $\Sigma^4$ and determine the total 
deflection angle $\alpha$ on the curved $(r, \phi)$ subspace 
${\cal M}^{\rm sub}$.

Here, let us name the curve $C_{\Gamma}$, e.g., Eq. (\ref{eq:Gamma2}),
``the unperturbed reference line'' which is the vertical projection of
null geodesic $\Gamma$ onto the curved $(r, \phi)$ subspace ${\cal M}^{\rm sub}$. 
The reason is that it is originally a null geodesic $\Gamma$ on 
the Minkowski $(\rho, \varphi)$ flat space (the background space) and 
it does not change its form on the curved $(r, \phi)$ subspace 
${\cal M}^{\rm sub}$ by vertical projection; therefore it seems to be 
reliable to regarded $C_{\Gamma}$ as the reference with respect to 
the actual null geodesic $\gamma$ on the curved $(r, \phi)$ subspace 
${\cal M}^{\rm sub}$
\footnote{
The authors of \cite{gibbons_werner2008a,ishihara_etal2016}
employed the circular arc with a constant coordinate radius 
as the curve $\gamma_P$ \cite{gibbons_werner2008a}, and 
$C_r$ \cite{ishihara_etal2016} instead of $C_{\Gamma}$ in our case.
However this seems to be inadequate because the circular arc is not 
related to the null geodesic in the background spacetime.}.

\begin{figure}[htbp]
\begin{center}
 \includegraphics[scale=0.3]{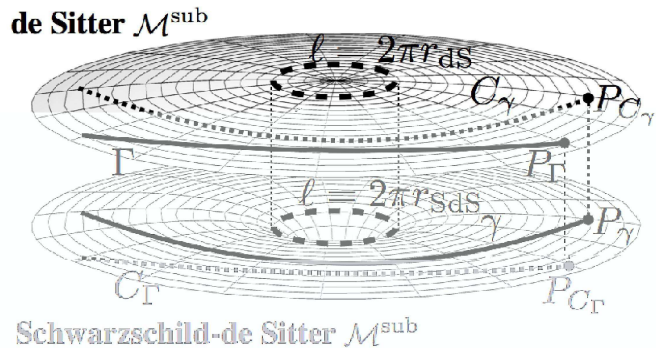}
 \caption{The relation between the de Sitter curved 
 $(r_{\rm dS}, \phi_{\rm dS})$ subspace ${\cal M}^{\rm sub}$ and 
 the Schwarzschild--de Sitter curved $(r_{\rm SdS}, \phi_{\rm SdS})$ 
 subspace ${\cal M}^{\rm sub}$. As in Fig. \ref{fig:arakida-fig2}, 
 the null geodesics $\Gamma, \gamma$ are represented by solid lines,  
 the curves $C_{\Gamma}, C_{\gamma}$ are represented by dotted lines, and
 the two circumferences $\ell$ are represented by dashed lines.
 The points $P_{\Gamma}$ on $\Gamma$ and $P_{C_{\Gamma}}$ 
 on $C_{\Gamma}$ have the same coordinate values and 
 the points $P_{\gamma}$ on $\gamma$ and $P_{C_{\gamma}}$ 
 on $C_{\gamma}$ also have the same coordinate values
 because the circumference radius 
 $\ell = 2\pi r_{\rm SdS} = 2\pi r_{\rm dS}$ and the vertical projection.
 \label{fig:arakida-fig3}}
\end{center}
\end{figure}

Next, in the case of the Schwarzschild--de Sitter spacetime, the spacetime
does not become asymptotically flat, and so the Minkowski
$(\rho, \varphi)$ flat space cannot be used as the background spacetime.
Hence, we assume the de Sitter spacetime to be the background spacetime
of the Schwarzschild--de Sitter spacetime, see Fig. \ref{fig:arakida-fig3}. 
In fact we can think of the Schwarzschild--de Sitter spacetime as 
the de Sitter spacetime distorted by the central mass $m$ 
(See section 14.4 in \cite{rindler2006}). From the considerations in
Appendix \ref{append:space}, the coordinate value of the curved $(r, \phi)$ 
subspace of the Schwarzschild--de Sitter space is the same as that of 
the de Sitter spacetime
\begin{eqnarray}
 r_{\rm SdS} = r_{\rm dS},\quad \phi_{\rm SdS} = \phi_{\rm dS}.
  \label{eq:gamma3}
\end{eqnarray}
because $r_{\rm dS}$ and $r{\rm SdS}$ are determined as the circumference radius
$\ell = 2\pi r_{\rm SdS} = 2\pi r_{\rm dS}$.
Furthermore, as will be seen later, the form of the null geodesic $\Gamma$
on the de Sitter space is similar to that on the Minkowski
$(\rho, \varphi)$ flat space; see Eqs. (\ref{eq:dSdiffeq1}) and 
(\ref{eq:dSdiffeq2}).
Therefore, in the Schwarzschild--de Sitter case, we adopt Eq. 
(\ref{eq:dSdiffeq2}) as the unperturbed reference line $C_{\Gamma}$
which is vertically projected from the de Sitter space to 
the Schwarzschild--de Sitter space.

So far, we have discussed the relation between the geodesic and the curve
on the curved $(r, \phi)$ subspaces ${\cal M}^{\rm sub}$
and its background space such as the Minkowski $(\rho, \varphi)$
flat space and the de Sitter subspace. However, in the present paper, 
we will be working in the optical reference geometry ${\cal M}^{\rm opt}$ 
to consider the behavior of the light ray and to apply the Gauss--Bonnet
theorem. As observed in section \ref{sec:optical}, the null geodesic 
(in our case, $\gamma$, $\gamma_S$ and $\gamma_R$; see Fig. 
\ref{fig:arakida-fig4} below) can be written in the same form both on 
the optical reference geometry ${\cal M}^{\rm opt}$ and the curved
$(r, \phi)$ subspace ${\cal M}^{\rm sub}$ due to the property of conformal
transformation. Moreover, the conformal transformation preserves the angle
between two vectors such as between $\gamma$ and $\gamma_S$ 
(namely $\varepsilon_S$), $C_{\Gamma}$ and $\gamma_S$ 
(namely $E_S$) and so on; see also Fig. \ref{fig:arakida-fig4}.
It is sufficient then to deduce the unperturbed reference line $C_{\Gamma}$ 
does not change its form both on the curved $(r, \phi)$ subspace 
${\cal M}^{\rm sub}$ and the optical reference geometry ${\cal M}^{\rm opt}$ 
as well as the null geodesics $\gamma$, $\gamma_S$ and $\gamma_R$.
Therefore in this paper, we take the stance that the unperturbed reference 
line $C_{\Gamma}$ on the optical reference geometry ${\cal M}^{\rm opt}$ 
obeys the same equation on the curved $(r, \phi)$ subspace ${\cal M}^{\rm sub}$.
\subsection{Total Deflection Angle}
\begin{figure}[htbp]
\begin{center}
 \includegraphics[scale=0.2]{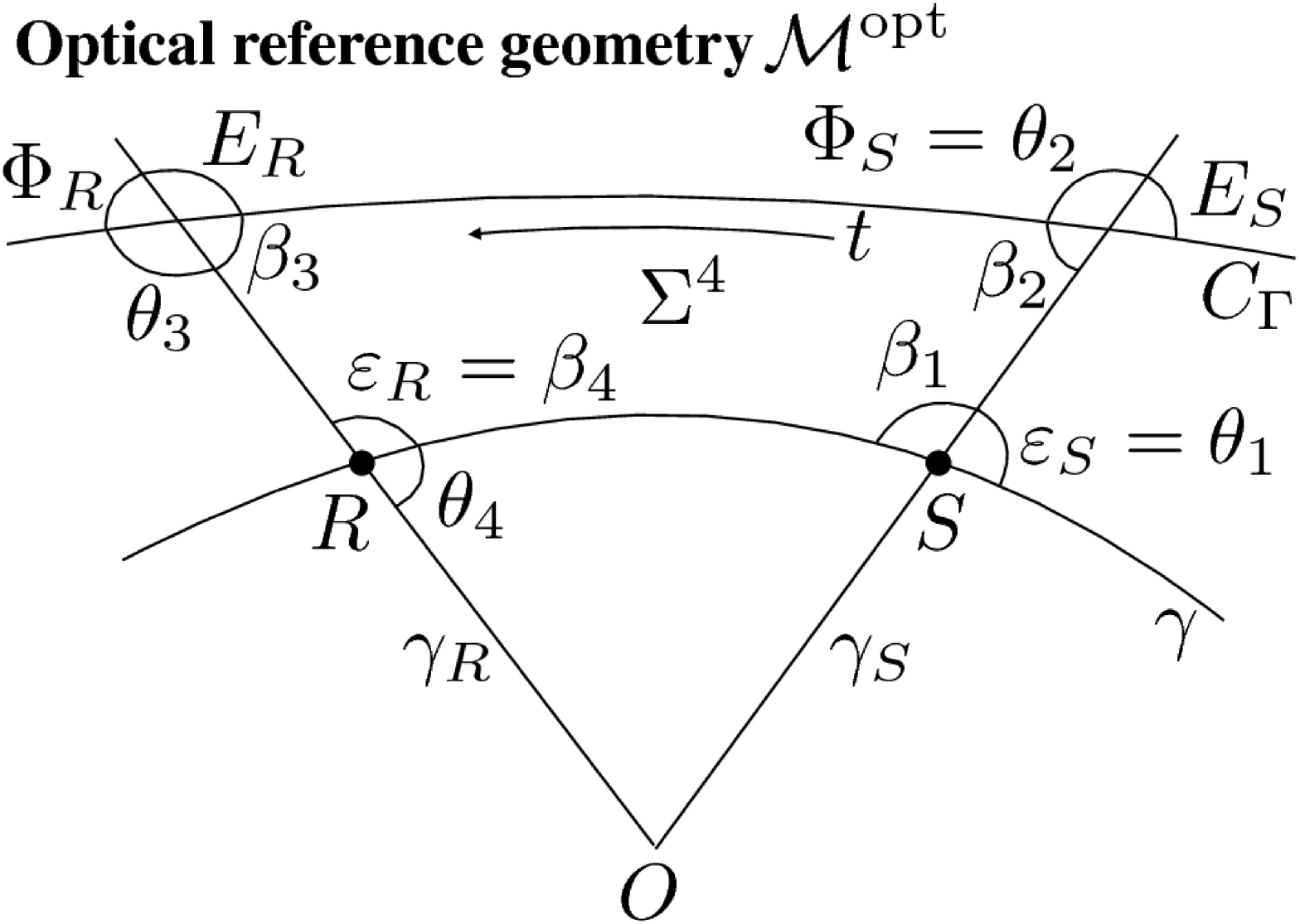}
 \caption{Schematic diagram of a quadrilateral $\Sigma^4$ on
 the optical reference geometry ${\cal M}^{\rm opt}$.
 A quadrilateral $\Sigma^4$ is bounded by the null geodesics $\gamma$,
 $\gamma_S$, $\gamma_R$ and the unperturbed reference 
 line $C_{\Gamma}$. The internal angles and the external angles 
 are $\beta_p$ and $\theta_p$ $(p = 1, 2, 3, 4)$, respectively.
 $\varepsilon_S$ and $\varepsilon_R$ are the angles between
 the geodesics $\gamma$ and $\gamma_S$ at $S$ and the angles between 
 $\gamma$ and $\gamma_R$ at $R$, respectively. $E_S$ and $E_R$ are 
 the angles between the unperturbed reference line $C_{\Gamma}$ and 
 the geodesic $\gamma_S$ and the angles between the unperturbed reference line 
 $C_{\Gamma}$ and $\gamma_R$, respectively. $\Phi_S$ and $\Phi_R$ were
 introduced in Eq. (\ref{eq:Schangle9}).
 \label{fig:arakida-fig4}}
\end{center}
\end{figure}

Let us define the total deflection angle. To this end, we construct 
a quadrilateral $\Sigma^4$ on the optical reference geometry 
${\cal M}^{\rm opt}$ that is bounded by the null geodesic 
of light ray $\gamma$, two radial null geodesics $\gamma_S$ connecting 
the center $O$ and the source $S$ of light ray and $\gamma_R$ connecting 
the center $O$ and the receiver $R$ of light ray, and the unperturbed 
reference line $C_{\Gamma}$ introduced in the previous 
subsection \ref{sec:rsl}; see Fig. \ref{fig:arakida-fig4}.

Let the internal angles of a quadrilateral $\Sigma^4$ be
$\beta_p ~ (p = 1, 2, 3, 4)$.
According to the value of the Gaussian curvature $K$, the sum of 
the internal angles $\beta_p$ of a quadrilateral $\Sigma^4$
on ${\cal M}^{\rm opt}$ becomes
\begin{eqnarray}
 \sum_{p = 1}^4 \beta_p
  \left\{
   \begin{array}{cl}
    > 2\pi & \mbox{for} ~K > 0\\
    = 2\pi & \mbox{for} ~K = 0\\
    < 2\pi & \mbox{for} ~K < 0\\
   \end{array}
   \right..
\end{eqnarray}
Then, let us suppose the total deflection angle $\alpha$ takes the positive
value, and we define the total deflection angle $\alpha$ as follows
\begin{eqnarray}
 \alpha \equiv
  \left|
   \sum_{p = 1}^4 \beta_p - 2\pi
  \right|.
\label{eq:angle1}
\end{eqnarray}
Next, we introduce four angles $\varepsilon_S$, $\varepsilon_R$,
$E_S$, and $E_R$ which pertain to the quadrilateral $\Sigma^4$
on ${\cal M}^{\rm opt}$. $\varepsilon_S$ and $\varepsilon_R$ are
angles between the trajectory of the light ray $\gamma$ and 
the radial null geodesics $\gamma_S$ and $\gamma_R$, respectively.
$E_S$ and $E_R$ are angles between the unperturbed reference
line $C_{\Gamma}$ and the radial null geodesics $\gamma_S$ and
$\gamma_R$, respectively.
We select these four angles in such a way that $\varepsilon_S$ and $E_S$
are corresponding angles, and as are $\varepsilon_R$ and $E_R$,
see Fig. \ref{fig:arakida-fig4}. We note that $\varepsilon_S$ and $E_S$ 
are a function of the angular coordinate value $\phi_S$, 
and $\varepsilon_R$ and $E_R$ are a function of $\phi_R$. 
The internal angles $\beta_i$ are then expressed as follows:
\begin{eqnarray}
   \beta_1 = \pi - \varepsilon_S,\quad
  \beta_2 = E_S,\quad
  \beta_3 = \pi - E_R,\quad
   \beta_4 = \varepsilon_R.
  \label{eq:angle2}
\end{eqnarray}
Therefore, from Eqs (\ref{eq:angle1}) and (\ref{eq:angle2}),
the total deflection angle $\alpha$ can be rewritten by as
\begin{eqnarray}
 \alpha =
  \left|(\varepsilon_R - E_R) - (\varepsilon_S - E_S)\right|.
  \label{eq:angle3}
\end{eqnarray}
Because we are working in the optical reference geometry
${\cal M}^{\rm opt}$ described by the optical metric $\bar{g}_{ij}$,
the angles $\varepsilon_S$ and $\varepsilon_R$ are computed by
using Eq. (\ref{eq:optical1}) as
\begin{eqnarray}
 \tan\varepsilon =
  \frac{\sqrt{\bar{g}_{\phi\phi}(r_{\gamma})}}
  {\sqrt{\bar{g}_{rr}(r_{\gamma})}}
  \left.\frac{d\phi}{dr}\right|_{\gamma}
  =
  \sqrt{f(r_{\gamma})}r_{\gamma}
  \left.\frac{d\phi}{dr}\right|_{\gamma},
  \label{eq:angle4}
\end{eqnarray}
where subscript $\gamma$ means $r$ and $d\phi/dr$ are the trajectory of
the light ray $\gamma$ and its differential equation, respectively,
and the optical metric $\bar{g}_{ij}$ is a function of $r_{\gamma}$.
On the other hand, $E_S$ and $E_R$ are similarly obtained as
\begin{eqnarray}
 \tan E = \frac{\sqrt{\bar{g}_{\phi\phi}(r_{C_{\Gamma}})}}
  {\sqrt{\bar{g}_{rr}(r_{C_{\Gamma}})}}
  \left.\frac{d\phi}{dr}\right|_{C_{\Gamma}}
  = \sqrt{f(r_{C_{\Gamma}})}r_{C_{\Gamma}}
  \left.\frac{d\phi}{dr}\right|_{C_{\Gamma}},
  \label{eq:angle5}
\end{eqnarray}
in which subscript $C_{\Gamma}$ indicates $r$ and $d\phi/dr$ are 
the unperturbed reference line $C_{\Gamma}$ and 
its differential equation, respectively. In this case, 
the optical metric $\bar{g}_{ij}$ is a function of $r_{C_{\Gamma}}$
\footnote{
The angle $\Psi$ in \cite{ishihara_etal2016} should be considered as 
$E$ instead of $\varepsilon$ in our case because the authors of
 \cite{ishihara_etal2016} seems to regard $1/r = \sin\phi/b$ as 
the equation of the light ray and inserted it into e.g., Eq. (32) 
in \cite{ishihara_etal2016} to calculate $\Psi$.
See also FIG. 4 in \cite{ishihara_etal2016}.
}.
From the practical point of view, the angles $\varepsilon$ and
$E$ should be computed by the tangent formula as 
Eqs. (\ref{eq:angle4}) and (\ref{eq:angle5}) instead of the sine and 
cosine formulas; see Appendix \ref{append:angle}.

Finally, let us relate the definition of the total deflection angle,
Eq. (\ref{eq:angle1}), to the Gauss--Bonnet theorem, Eq. (\ref{eq:GB1}).
For this purpose, we prepare the external angles $\theta_p$
in such a way that
\begin{eqnarray}
   \theta_1 = \varepsilon_S,\quad
  \theta_2 = \pi - E_S,\quad
  \theta_3 = E_R,\quad
  \theta_4 = \pi - \varepsilon_R.
  \label{eq:angle6}
\end{eqnarray}
It should be emphasized here that due to the conformal transformation,
all the angles, $\beta_p$, $\theta_p$, $\varepsilon_S$, $\varepsilon_R$, 
$E_S$ and $E_R$, are given the same value both on the optical reference
geometry ${\cal M}^{\rm opt}$ and the curved $(r, \phi)$ subspace
${\cal M}^{\rm sub}$. The sum of the external angles becomes
\begin{eqnarray}
 \sum_{p = 1}^4 \theta_p = 2\pi + (\varepsilon_S - E_S)
  - (\varepsilon_R - E_R).
  \label{eq:angle7}
\end{eqnarray}
Because $\gamma$, $\gamma_S$ and $\gamma_R$ are null geodesics,
it is enough to integrate the line integral of the geodesic curvature 
$\kappa_g$ along the curve $C_{\Gamma}$, then, according to 
Eqs. (\ref{eq:GB1}), (\ref{eq:angle3}) and (\ref{eq:angle7}), we find,
\begin{eqnarray}
 \alpha &=&
  \left|(\varepsilon_R - E_R) - (\varepsilon_S - E_S)\right|
  \label{eq:angle8}
  \\
 &=&
 \left|
  \iint_{\Sigma^4}Kd\sigma + \int_{C_{\Gamma}}\kappa_g dt
	\right|,
 \label{eq:angle9}
\end{eqnarray}
noting that the Gaussian curvature $K$ and the geodesic curvature $\kappa_g$
are determined by the optical metric $\bar{g}_{ij}$ instead of $g_{ij}$.

Let us call Eq. (\ref{eq:angle8}) {\it the angular formula} of
the total deflection angle $\alpha$ and Eq. (\ref{eq:angle9}) 
{\it the integral formula} of the total deflection angle $\alpha$.
It should be pointed out that due to the conformal transformation,
the angular formula, Eq. (\ref{eq:angle8}), holds not only on 
the optical reference geometry ${\cal M}^{\rm opt}$ 
but also on the curved $(r, \phi)$ subspace ${\cal M}^{\rm sub}$
\footnote{
It is clear that the one-sided deflection angle 
$\epsilon = \psi - \phi$ in \cite{rindler2007} is rewritten as
$\epsilon = \varepsilon - \phi$ in our notation. 
This should then be corrected as $\epsilon = \psi - E$, replacing 
the angular coordinate value $\phi$ by $E$.
Our one-sided deflection angle $\epsilon = \varepsilon - E$ seems to be
equivalent to Eq. (140), $\beta_M = \beta_{M2} - \beta_{M1}$, in
\cite{lebedevlake2013}.}. 
The angular formula, Eq. (\ref{eq:angle8}) is directly related to
the definition of the total deflection angle, Eq. (\ref{eq:angle1}), 
whose meaning is geometrically clear.
By contrast, in the case of the integral formula, the results of the
integration differ for the optical reference geometry ${\cal M}^{\rm opt}$
and the curved $(r, \phi)$ subspace ${\cal M}^{\rm sub}$ because
the conformal transformation rescales the coordinate values
under the transformation $g_{ij} \leftrightarrow \bar{g}_{ij}$.
\section{Asymptotically Flat Spacetime\label{sec:asymptotic}}
In the beginning, let us consider the asymptotically flat spacetime, 
namely the Schwarzschild spacetime. We apply the formulas Eqs. (\ref{eq:angle8}) 
and (\ref{eq:angle9}) to calculate the total deflection angle $\alpha$, 
and show that the two formulas give the same results.
\subsection{Schwarzschild Spacetime}
In the case of the Schwarzschild spacetime, the form of $f(r)$ is
\begin{eqnarray}
 f_{\rm Sch}(r) = 1 - \frac{2m}{r},
  \label{eq:fSch}
\end{eqnarray}
in which $m$ is the mass of the central object. We assume that
the source $S$ and the receiver $R$ of light rays are located at
a finite distance from the center on the curved region of
${\cal M}^{\rm opt}$, and we obtain the total deflection angle up to
the order ${\cal O}(m^2)$.
\subsubsection{Angular Formula}
First, we calculate the total deflection angle $\alpha$ by using
angular formula, Eq. (\ref{eq:angle8}). $\varepsilon_S$
and $\varepsilon_R$ are obtained by Eq. (\ref{eq:angle4}):
\begin{eqnarray}
 \tan\varepsilon
  = \sqrt{1 - \frac{2m}{r_{\gamma}}}r_{\gamma}
  \left.\frac{d\phi}{dr}\right|_{\gamma}
 = \sqrt{1 - \frac{2m}{r_{\gamma}}}\frac{b}{r_{\gamma}}
      \left(
       1 + \frac{2mb^2}{r^3_{\gamma}} - \frac{b^2}{r^2_{\gamma}}
	\right)^{-\frac{1}{2}},
 \label{eq:Schangle1}
\end{eqnarray}
where we used the differential equation of the light ray $\gamma$
on ${\cal M}^{\rm opt}$:
\begin{eqnarray}
 \left.\left(\frac{dr}{d\phi}\right)^2\right|_{\gamma}
  = r^2_{\gamma}
  \left(
   \frac{r^2_{\gamma}}{b^2} + \frac{2m}{r_{\gamma}} - 1
  \right).
  \label{eq:Schdiffeq1}
\end{eqnarray}
From Eq. (\ref{eq:Schdiffeq1}), the trajectory of the light ray $\gamma$
is given by up to the order ${\cal O}(m^2)$; see Eq. (7) in \cite{ak2012}:
\begin{eqnarray}
 \frac{1}{r_{\gamma}} &=& \frac{\sin\phi}{b} + \frac{m}{2b^2}(3 + \cos 2\phi)
  \nonumber\\
  & &+ \frac{m^2}{16b^3}
  \left[37\sin\phi + 30(\pi - 2\phi)\cos\phi - 3\sin 3\phi\right]
  + {\cal O}(m^3).
  \label{eq:Schtrajectory1}
\end{eqnarray}
Inserting Eq. (\ref{eq:Schtrajectory1}) into Eq. (\ref{eq:Schangle1}) and 
expanding up to the order ${\cal O}(m^2)$, $\tan\varepsilon$ is written as
\begin{eqnarray}
 \tan\varepsilon
  = \tan\phi + \frac{2m}{b}\sec\phi +
  \frac{m^2}{8b^2}
  \left[15\sec^2\phi (\pi - 2\phi + \sin 2\phi)\right]
  + {\cal O}(m^3).
   \label{eq:Schangle2}
\end{eqnarray}
Thus, $\varepsilon$ is
\begin{eqnarray}
 \varepsilon &=&
  \arctan\left\{
	  \tan\phi + \frac{2m}{b}\sec\phi +
	  \frac{m^2}{8b^2}
	  [15\sec^2\phi (\pi - 2\phi + \sin 2\phi)] 
	 \right\}\nonumber\\
 &=&
  \phi + \frac{2m}{b}\cos\phi +
  \frac{m^2}{8b^2}
  \left[15(\pi - 2\phi) - \sin 2\phi\right] + {\cal O}(m^3).
  \label{eq:Schangle3}
\end{eqnarray}
Angle $E$ is obtained by using Eq. (\ref{eq:angle5}):
\begin{eqnarray}
 \tan E = \sqrt{1 - \frac{2m}{r_{C_{\Gamma}}}} r_{C_{\Gamma}}
	     \left.\frac{d\phi}{dr}\right|_{C_{\Gamma}}
 = \sqrt{1 - \frac{2m}{r_{C_{\Gamma}}}}\tan \phi,
  \label{eq:Schangle4}
\end{eqnarray}
which is obtained from the relation
\begin{eqnarray}
 \left.\frac{1}{r_{C_{\Gamma}}^2}
  \left(\frac{dr}{d\phi}\right)^2\right|_{C_{\Gamma}}
  = \left(\frac{r^2_{C_{\Gamma}}}{b^2} - 1\right)
  = \frac{1}{\tan^2\phi},
  \label{eq:Schdiffeq2}
\end{eqnarray}
noting that $1/r_{C_{\Gamma}} = \sin\phi/b$.
In a similar way to the derivation of Eq. (\ref{eq:Schangle3}), $E$ becomes
\begin{eqnarray}
 E = \phi - \frac{m}{b}\sin^2\phi\cos\phi
  - \frac{m^2}{2b^2}\sin^3\phi\cos\phi (2 - \cos 2\phi)
  + {\cal O}(m^3).
  \label{eq:Schangle5}
\end{eqnarray}
Therefore, from Eqs. (\ref{eq:Schangle3}), (\ref{eq:Schangle5}),
and (\ref{eq:angle8}), the total deflection angle in the Schwarzschild
spacetime is obtained as
\begin{eqnarray}
 \alpha_{\rm Sch} &=& (\varepsilon_S - E_S) - (\varepsilon_R - E_R)
  \nonumber\\
  &=&
  - \frac{m}{b}
  \left[
   2(\cos\phi_R - \cos\phi_S)
   + \sin^2\phi_R\cos\phi_R - \sin^2\phi_S\cos\phi_S
  \right]\nonumber\\
  & &+ \frac{m^2}{8b^2}
  \left\{
   30(\phi_R - \phi_S) + \sin 2\phi_R - \sin 2\phi_S
   \right.\nonumber\\
   & &\left.-4\left[
      \sin^3\phi_R\cos\phi_R (2 - \cos 2\phi_R)
      -
      \sin^3\phi_S\cos\phi_S (2 - \cos 2\phi_S)
     \right]
      \right\}\nonumber\\
 & &
   + {\cal O}(m^3),
   \label{eq:Schangle6}
\end{eqnarray}
where the sign is taken in such a sense that the total deflection 
angle $\alpha$ is positive.
\subsubsection{Integral Formula}
Next, let us show that the integral formula, Eq. (\ref{eq:angle9}),
also gives the same result as Eq. (\ref{eq:Schangle6}).
At first, we integrate the areal integral of the Gaussian
curvature $K^{\rm Sch}$. On the basis of the optical metric 
Eq. (\ref{eq:optical1}) and Eq. (\ref{eq:fSch}), the Gaussian
curvature Eq. (\ref{eq:GB2}) is given by
\begin{eqnarray}
 K^{\rm Sch} = - \frac{2m}{r^3}\left(1 - \frac{3m}{2r}\right),
  \label{eq:KSch}
\end{eqnarray}
and the areal element $d\sigma = \sqrt{\det|\bar{g}|}drd\phi$ is
\begin{eqnarray}
 d\sigma = r\left(1 - \frac{2m}{r}\right)^{-\frac{3}{2}}drd\phi.
  \label{eq:dSSch}
\end{eqnarray}
The quadrilateral $\Sigma^4$ is bounded by the geodesic of the light ray 
$\gamma$, the two radial null geodesics $\gamma_S$ and $\gamma_R$ and 
the unperturbed reference line $C_{\Gamma}$. $\gamma_S$ and $\gamma_R$ 
are parametrized by the angular coordinate values $\phi_S$ and $\phi_R$, 
respectively. Hence, using Eqs (\ref{eq:KSch}) and (\ref{eq:dSSch}), and
expanding up to the order ${\cal O}(m^2)$, we obtain
\begin{eqnarray}
 \iint_{\Sigma^4}K^{\rm Sch}d\sigma
  &=& -\int_{\phi_S}^{\phi_R}\int_{r_{\gamma}}^{r_{C_{\Gamma}}}
  \left(
   \frac{2m}{r^2} + \frac{3m^2}{r^3} 
		\right)drd\phi + {\cal O}(m^3)
  \nonumber\\
 &=&
  - \frac{m^2}{8b^2}
  \left[
   24(\phi_R - \phi_S) + 4(\sin 2\phi_R - \sin 2\phi_S)
	 \right]
  + {\cal O}(m^3),
  \label{eq:Schangle7}
\end{eqnarray}
where $1/r_{C_{\Gamma}} = \sin\phi/b$ and Eq. (\ref{eq:Schtrajectory1}) 
is used as the trajectory of the light ray $r_{\gamma}$. However in practice, 
it is enough to insert the order ${\cal O}(m)$ solution of $r_{\gamma}$ to 
integrate the areal integral of $K^{\rm Sch}$ because the leading order of 
the Gaussian curvature $K^{\rm Sch}$ is ${\cal O}(m)$.

Subsequently, we calculate the line integral of the geodesic curvature
$\kappa_g^{\rm Sch}$. The geodesic curvature Eq. (\ref{eq:GB3}) is expressed as
\begin{eqnarray}
  \kappa_g^{\rm Sch} = \frac{b}{r^2}\left(1 - \frac{3m}{r}\right)
   \sqrt{1 - \frac{2m}{r}} + \frac{d\Phi}{dt}.
   \label{eq:kgSch}
\end{eqnarray}
Noting that the integration of $\kappa_g$ is carried out along the unperturbed 
reference line $1/r_{C_{\Gamma}} = \sin\phi/b$,
\begin{eqnarray}
 \int_{C_{\Gamma}} \kappa_g^{\rm Sch}dt
  &=&
  \int_{\phi_S}^{\phi_R}
  \left(1 - \frac{2m}{r_{C_{\Gamma}}} - \frac{3m^2}{2r^2_{C_{\Gamma}}}\right)d\phi
  + \int_S^R \frac{d\Phi}{dt}dt + {\cal O}(m^3)
  \nonumber\\
 &=&
  \phi_R - \phi_S + \frac{2m}{b}(\cos\phi_R - \cos\phi_S)
  \nonumber\\
  & &- \frac{m^2}{8b^2}
  \left[6(\phi_R - \phi_S) - 3(\sin 2\phi_R - \sin 2\phi_S)\right]
  \nonumber\\
  & &+ \Phi_R - \Phi_S + {\cal O}(m^3)
  \label{eq:Schangle8}
\end{eqnarray}
where we changed the integration variable of the first term from $t$ to $\phi$ using
\begin{eqnarray}
  dt = \frac{r^2}{b} \left(1 - \frac{2m}{r}\right)^{-1}d\phi,
  \label{eq:dtSch}
\end{eqnarray}
which is derived from Eq. (\ref{eq:dpdt}). 
Because the arc length parameter $t$ moves along the curve $C_{\Gamma}$
in the counterclockwise direction, the direction of the tangent vector of 
curve $C_{\Gamma}$ is also counterclockwise; see Fig. \ref{fig:arakida-fig4}.
Then $\Phi_S$ and $\Phi_R$ should be given by
\begin{eqnarray}
 \Phi = \pi - E.
  \label{eq:Schangle9}
\end{eqnarray}
Substituting Eqs. (\ref{eq:Schangle5}) and (\ref{eq:Schangle9})
into (\ref{eq:Schangle8}), we have
\begin{eqnarray}
  \int_{C_{\Gamma}} \kappa_g^{\rm Sch}dt
   &=&
   \frac{m}{b}
   \left[
    2(\cos\phi_R - \cos\phi_S)
    + \sin^2\phi_R\cos\phi_R - \sin^2\phi_S\cos\phi_S
	    \right]\nonumber\\
 & &
  - \frac{m^2}{8b^2}
  \left\{
   6(\phi_R - \phi_S) - 3(\sin 2\phi_R - \sin 2\phi_S)
     \right.\nonumber\\
 & &
   \left. - 4\left[
       \sin^3\phi_R\cos\phi_R(2 - \cos 2\phi_R)
       -
       \sin^3\phi_S\cos\phi_S(2 - \cos 2\phi_S)
      \right]
		  \right\}
   \nonumber\\
 & &+ {\cal O}(m^3)
  \label{eq:Schangle10}
\end{eqnarray}
From Eqs. (\ref{eq:Schangle7}), (\ref{eq:Schangle10}) and
(\ref{eq:angle9}), the total deflection angle $\alpha$ becomes
\begin{eqnarray}
 \alpha_{\rm Sch} &=& -\iint_{\Sigma^4}K^{\rm Sch}d\sigma
  - \int_{C_{\Gamma}} \kappa_g^{\rm Sch}dt
  \nonumber\\
 &=&
    - \frac{m}{b}
   \left[
    2(\cos\phi_R - \cos\phi_S)
    + \sin^2\phi_R\cos\phi_R - \sin^2\phi_S\cos\phi_S
	    \right]\nonumber\\
 & &
  + \frac{m^2}{8b^2}
  \left\{
   30(\phi_R - \phi_S) + \sin 2\phi_R - \sin 2\phi_S
     \right.\nonumber\\
 & &
   \left. -4\left[
       \sin^3\phi_R\cos\phi_R(2 - \cos 2\phi_R)
       -
       \sin^3\phi_S\cos\phi_S(2 - \cos 2\phi_S)
      \right]
  \right\}\nonumber\\ 
   & &+ {\cal O}(m^3).
   \label{eq:Schangle11}
\end{eqnarray}
We find that Eq. (\ref{eq:Schangle11}) is in perfect agreement 
with Eq. (\ref{eq:Schangle6}).
\subsection{Limit of Infinite Distance Source and Receiver}
Let us confirm that Eqs (\ref{eq:Schangle6}) and (\ref{eq:Schangle11})
coincide with Epstein--Shapiro's formula \cite{epstein1980} in the limit 
of the infinite distance source $S$ and receiver $R$. To this end, 
we put $\phi_R \rightarrow \pi$ and $\phi_S \rightarrow 0$; then
\begin{eqnarray}
 \alpha_{\rm Sch} \rightarrow
  \frac{4m}{b} + \frac{15\pi m^2}{4b^2} +
  {\cal O}(m^3).
  \label{eq:Schangle12}
\end{eqnarray}
Two points are noteworthy; first, we observe the breakdown of 
Eq. (\ref{eq:Schangle12}) in terms of the integral formula,
Eq. (\ref{eq:angle9}). The areal integral part is
\begin{eqnarray}
 -\iint_{\Sigma^4}K^{\rm Sch}d\sigma
  = \frac{12\pi m^2}{4b^2}.
  \label{eq:Schangle13}
\end{eqnarray}
Thus, the areal integral does not contribute to the first-order
in $m$ because the leading order of the Gaussian curvature is
${\cal O}(m)$, which is enough to insert the zero-th order
trajectory of light rays $1/r_{\gamma} = \sin\phi/b$ when calculating
the total deflection angle $\alpha$ up to the order ${\cal O}(m)$.
As a consequence, the upper limit $r_{C_{\Gamma}}$ of integration
in $r$ coincides with the lower limit $r_{\gamma}$, and
the areal integral vanishes.
On the other hand, the line integral part becomes
\begin{eqnarray}
 -\int_{C_{\Gamma}}\kappa_g^{\rm Sch}dt
  = \frac{4m}{b} + \frac{3\pi m^2}{4b^2}.
  \label{eq:Schangle14}
\end{eqnarray}

Second, in terms of the angular formula, Eq. (\ref{eq:angle8}),
it is found for $\phi_S \rightarrow 0, \phi_R \rightarrow \pi$ that
$E_S \rightarrow 0, E_R \rightarrow \pi$; see Eq. (\ref{eq:Schangle5}).
Hence:
\begin{eqnarray}
 \alpha \rightarrow
  \varepsilon_S (\phi_S \rightarrow 0) -
  \varepsilon_R (\phi_R \rightarrow \pi) + \pi,
  \label{eq:Schangle15}
\end{eqnarray}
which has a different sign for $\pi$ compared to the standard
treatment found in textbooks and the literature. 
This difference is due to the fact that in the standard treatment,
the total deflection angle is calculated on the Minkowskian
flat space then 
$\varepsilon_R (\phi_R \rightarrow \pi) - \varepsilon_S (\phi_S \rightarrow 0) > \pi$, 
and $\alpha \rightarrow \varepsilon_R (\phi_R \rightarrow \pi) - 
\varepsilon_S (\phi_S \rightarrow 0) - \pi$; whereas, in the optical
reference geometry ${\cal M}^{\rm opt}$ or the curved $(r, \phi)$
subspace ${\cal M}^{\rm sub}$, the Gaussian curvature $K^{\rm Sch}$
takes the negative value $K^{\rm Sch} < 0$ so that the total deflection
angle is given by Eq. (\ref{eq:Schangle15}).
\section{Non-asymptotically Flat Spacetime\label{sec:non-asymptotic}}
The goal of this section is to investigate and reveal the contribution
of the cosmological constant $\Lambda$ to the bending of light
in terms of the Schwarzschild--de Sitter spacetime.
As is the case with the Schwarzschild spacetime in the previous section,
we demonstrate that the angular formula Eq. (\ref{eq:angle8}) and
the integral formula Eq. (\ref{eq:angle9}) lead to the same result.
\subsection{de Sitter Spacetime\label{sec:desitter}}
Before discussing the total deflection angle in the Schwarzschild--de Sitter
spacetime, let us examine the light rays in the de Sitter spacetime
which can be considered as the background spacetime of the Schwarzschild--de
Sitter spacetime. The conclusion arrived at here may be trivial; 
however it provides an important suggestion when considering 
the total deflection angle in the Schwarzschild--de Sitter spacetime.

In the case of the de Sitter spacetime, $f(r)$ is given by
\begin{eqnarray}
 f_{\rm dS}(r) = 1 - \frac{\Lambda}{3}r^2,
  \label{eq:fdS}
\end{eqnarray}
where $\Lambda$ is the cosmological constant, and the differential
equation of the light ray $\gamma$ on the optical reference geometry
${\cal M}^{\rm opt}$ is
\begin{eqnarray}
 \left.\left(\frac{dr}{d\phi}\right)^2\right|_{\gamma} =
  r^2_{\gamma}
  \left[
   \left(\frac{1}{b^2} + \frac{\Lambda}{3}\right)r^2_{\gamma} - 1
  \right].
  \label{eq:dSdiffeq1}
\end{eqnarray}
If we reduce the two constants $1/b^2$ and $\Lambda/3$ to a single constant,
\begin{eqnarray}
 \frac{1}{B^2} \equiv \frac{1}{b^2} + \frac{\Lambda}{3},
  \label{eq:impact2}
\end{eqnarray}
and Eq. (\ref{eq:dSdiffeq1}) gives the similar form of equation as
Eq. (\ref{eq:Gamma2}):
\begin{eqnarray}
 \frac{1}{r_{\gamma}} = \frac{\sin\phi}{B}.
  \label{eq:dSdiffeq2}
\end{eqnarray}
We notice that from Eqs. (\ref{eq:dSdiffeq1}) and (\ref{eq:impact2}), 
$B$ can be expressed by $r_0$ which is the radial coordinate at the point 
of closest approach of light trajectory,
\begin{eqnarray}
 \frac{1}{B^2} = \frac{1}{b^2} + \frac{\Lambda}{3} = \frac{1}{r^2_0},
  \label{eq:impact3}
\end{eqnarray}
because $\left.dr/d\phi\right|_{r_0} = 0$. Then $B$ is obtained without 
any information of $\Lambda$. Accordingly, the equation of the light 
trajectory $\gamma$ corresponds to that of the unperturbed reference 
line $C_{\Gamma}$ (please imagine that in Fig. \ref{fig:arakida-fig5} below,
for $m \rightarrow 0$, the Schwarzschild--de Sitter space reduces to
de Sitter space); then $r_{\gamma} = r_{C_{\Gamma}}$, and 
conversely $C_{\Gamma}$ is now the null geodesic $\gamma$ itself. 
In this case, $\varepsilon_S = E_S$ and $\varepsilon_R = E_R$,
and $\gamma$ coincides with $C_{\Gamma}$, see Fig. \ref{fig:arakida-fig5}.
\begin{figure}[htbp]
\begin{center}
 \includegraphics[scale=0.25,angle=-90]{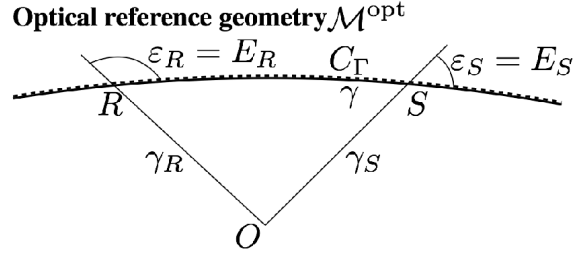}
 \caption{Relation between $\gamma$ (solid line) and 
 $C_{\Gamma}$ (dotted line) on de Sitter space. Because 
 $\gamma$ and $C_{\Gamma}$ coincide, the area of quadrilateral
 $\Sigma^4$ is zero.
 \label{fig:arakida-fig5}}
\end{center}
\end{figure}
From the angular formula, Eq. (\ref{eq:angle8}):
\begin{eqnarray}
 \alpha_{\rm dS} =
  \left|(\varepsilon_R - E_R) - (\varepsilon_S - E_S)\right| = 0.
  \label{eq:dSangle1}
\end{eqnarray}
Because $C_{\Gamma}$ is now the null geodesic $\gamma$, 
$\kappa_g^{\rm dS} = 0$, and we have
\footnote{From the result of Eq. (\ref{eq:dSangle2}), we may immediately 
say that the light ray in the de Sitter spacetime does not bend.}
\begin{eqnarray}
 \int_{C_{\Gamma}}\kappa_g^{\rm dS}dt
  =
  \int_{\gamma}\kappa_g^{\rm dS}dt
  = 0.
  \label{eq:dSangle2}
\end{eqnarray}
The Gaussian curvature of the de Sitter spacetime is
\begin{eqnarray}
 K^{\rm dS} = -\frac{\Lambda}{3} \ne 0,
  \label{eq:dSangle3}
\end{eqnarray}
but the upper limit $r_{C_{\Gamma}}$ and lower limit $r_{\gamma}$ of
integration in $r$ are the same $r_{\gamma} = r_{C_{\Gamma}}$
(see Fig. \ref{fig:arakida-fig5}); consequently:
\begin{eqnarray}
 \iint_{\Sigma^4}K^{\rm dS}d\sigma = 0.
  \label{eq:dSangle4}
\end{eqnarray}
Hence, we also obtain by means of the integral formula:
\begin{eqnarray}
 \alpha_{\rm dS} = -\iint_{\Sigma^4}K^{\rm dS}d\sigma -
  \int_{C_{\Gamma}}\kappa_g^{\rm dS}dt = 0.
  \label{eq:dSangle5}
\end{eqnarray}
We can conclude that in the de Sitter spacetime
the total deflection angle of the light ray is zero, 
$\alpha_{\rm dS} = 0$ then the light rays do not bend.
\subsection{Schwarzschild--de Sitter Spacetime}
From now on, we examine the total deflection angle $\alpha$ in
the Schwarzschild--de Sitter spacetime. The total deflection angle
is calculated up to the order ${\cal O}(m^2, \Lambda, \Lambda m)$.
However, as will be clarified later, the total deflection angle
does not contain ${\cal O}(\Lambda)$ terms though they appear
in the intermediate steps of the calculation.

The Schwarzschild--de Sitter/Kottler spacetime is characterized by
\cite{kottler1918}:
\begin{eqnarray}
 f_{\rm SdS}(r) = 1 - \frac{2m}{r} - \frac{\Lambda}{3}r^2.
  \label{eq:fSdS}
\end{eqnarray}
where $m$ is the mass of the central body and $\Lambda$ is the cosmological
constant. As in the case of the Schwarzschild spacetime, the total deflection
angle is calculated in such a way that the source $S$ and the receiver $R$ 
of the light ray are located at a finite distance from the center.
\subsubsection{Angular Formula}
The angle $\varepsilon$ is obtained from Eq. (\ref{eq:angle4}):
\begin{eqnarray}
 \tan\varepsilon
  = \sqrt{1 - \frac{2m}{r_{\gamma}} - \frac{\Lambda}{3}r^2_{\gamma}}
  r_{\gamma}\left.\frac{d\phi}{dr}\right|_{\gamma}
 = \sqrt{1 - \frac{2m}{r_{\gamma}} - \frac{\Lambda}{3}r^2_{\gamma}}
  \frac{B}{r_{\gamma}}
  \left(
   1 + \frac{2mB^2}{r^3_{\gamma}} - \frac{B^2}{r^2_{\gamma}}
  \right)^{-\frac{1}{2}},
  \label{eq:SdSangle1}
\end{eqnarray}
because the differential equation of the light ray in the
Schwarzschild--de Sitter spacetime is
\begin{eqnarray}
 \left.\left(\frac{dr}{d\phi}\right)^2\right|_{\gamma}
  = r^2_{\gamma}
  \left(
   \frac{r^2_{\gamma}}{B^2} + \frac{2m}{r_{\gamma}} - 1
  \right),
  \label{eq:SdSdiffeq1}
\end{eqnarray}
where we used Eq. (\ref{eq:impact2}). The equation of the
light trajectory $\gamma$ on the optical reference geometry
${\cal M}^{\rm opt}$ has the same form as Eq.
(\ref{eq:Schtrajectory1}):
\begin{eqnarray}
 \frac{1}{r_{\gamma}} &=& \frac{\sin\phi}{B} + \frac{m}{2B^2}(3 + \cos 2\phi)
  \nonumber\\
  & &+ \frac{m^2}{16B^3}
  \left[37\sin\phi + 30(\pi - 2\phi)\cos\phi - 3\sin 3\phi\right]
  + {\cal O}(m^3),
  \label{eq:SdStrajectory1}
\end{eqnarray}
except that the constant $b$ is replaced by $B$ (see Eq. (\ref{eq:impact2})). 
We mention that from Eqs. (\ref{eq:SdSdiffeq1}) and (\ref{eq:impact2}), 
$B$ can be written by $r_0$ being the radial coordinate at the point of 
closest approach of light trajectory
\begin{eqnarray}
 \frac{1}{B^2} = \frac{1}{b^2} + \frac{\Lambda}{3} 
  = \frac{1}{r^2_0} - \frac{2m}{r^3_0},
  \label{eq:impact4}
\end{eqnarray}
because of $\left.dr/d\phi\right|_{r_0} = 0$, then $B$ is obtained 
without any information of $\Lambda$.

Inserting Eq. (\ref{eq:SdStrajectory1}) into Eq. (\ref{eq:SdSangle1})
and expanding up to the order ${\cal O}(m^2, \Lambda, \Lambda m)$, one obtains,
\begin{eqnarray}
 \tan\varepsilon
  &=&
  \tan\phi + \frac{2m}{B}\sec\phi +
  \frac{m^2}{8B^2}
  [15\sec^2\phi (\pi - 2\phi + \sin 2\phi)]\nonumber\\
 & &
  - \frac{\Lambda B^2}{6}\csc\phi\sec\phi
  - \frac{\Lambda Bm}{3}(1 - \cot^2 \phi)\sec\phi
  + {\cal O}(m^3, \Lambda^2, \Lambda m^2).
   \label{eq:SdSangle2}
\end{eqnarray}
As with Eq. (\ref{eq:Schangle3}), $\varepsilon$ is given by
\begin{eqnarray}
 \varepsilon &=&
  \phi + \frac{2m}{B}\cos\phi +
  \frac{m^2}{8B^2}
  \left[15(\pi - 2\phi) - \sin 2\phi\right]
  \nonumber\\
  & &- \frac{\Lambda B^2}{6}\cot\phi
  + \frac{\Lambda Bm}{3}\cot\phi\csc\phi
+ {\cal O}(m^3, \Lambda^2, \Lambda m^2).
\label{eq:SdSangle3}
\end{eqnarray}
The first line of Eq (\ref{eq:SdSangle3}) corresponds to Eq. (\ref{eq:Schangle3}), 
with $b$ being replaced by $B$ in the Schwarzschild part, and the second line 
is due to the contribution of the cosmological constant $\Lambda$.

On the other hand, $E$ is calculated with Eq. (\ref{eq:angle5}):
\begin{eqnarray}
 \tan E = \sqrt{1 - \frac{2m}{r_{C_{\Gamma}}} -
  \frac{\Lambda}{3}r^2_{C_{\Gamma}}}
  r_{C_{\Gamma}}\left.\frac{d\phi}{dr}\right|_{C_{\Gamma}}
  =
   \sqrt{1 - \frac{2m}{r_{C_{\Gamma}}}
   - \frac{\Lambda}{3}r^2_{C_{\Gamma}}}\tan\phi,
  \label{eq:SdSangle4}
\end{eqnarray}
because $1/r_{C_{\Gamma}} = \sin\phi/B$; then
\begin{eqnarray}
 \left.\frac{1}{r_{C_{\Gamma}}^2}
  \left(\frac{dr}{d\phi}\right)^2\right|_{C_{\Gamma}}
  = \left(\frac{r^2_{C_{\Gamma}}}{B^2} - 1\right)
  = \frac{1}{\tan^2\phi}.
  \label{eq:SdSdiffeq2}
\end{eqnarray}
It follows that
\begin{eqnarray}
 E &=& 
 \phi - \frac{m}{B}\sin^2\phi\cos\phi
 - \frac{m^2}{2B^2}\sin^3\phi\cos\phi (2 - \cos 2\phi)
 \nonumber\\
 & &- \frac{\Lambda B^2}{6}\cot\phi
 - \frac{\Lambda Bm}{6}\cos\phi(2 - \cos 2\phi)
  + {\cal O}(m^3, \Lambda^2, \Lambda m^2).
  \label{eq:SdSangle5}
\end{eqnarray}
As in Eq. (\ref{eq:SdSangle3}), the first line of Eq. (\ref{eq:SdSangle5})
corresponds to Eq. (\ref{eq:Schangle5}), the Schwarzschild part with replaced 
$b$ by $B$, and the second line is due to the contribution of the cosmological
constant $\Lambda$.

As a result, the total deflection angle $\alpha$ becomes, using 
Eqs. (\ref{eq:SdSangle3}), (\ref{eq:SdSangle5}), and (\ref{eq:angle8}),
\begin{eqnarray}
 \alpha_{\rm SdS} &=& (\varepsilon_S - E_S) - (\varepsilon_R - E_R)
  \nonumber\\
  &=&
  - \frac{m}{B}
  \left[
   2(\cos\phi_R - \cos\phi_S)
   + \sin^2\phi_R\cos\phi_R - \sin^2\phi_S\cos\phi_S
  \right]\nonumber\\
  & &+ \frac{m^2}{8B^2}
  \left\{
   30(\phi_R - \phi_S) + \sin 2\phi_R - \sin 2\phi_S
   \right.\nonumber\\
   & &\left.-4\left[
      \sin^3\phi_R\cos\phi_R (2 - \cos 2\phi_R)
      -
      \sin^3\phi_S\cos\phi_S (2 - \cos 2\phi_S)
     \right]
      \right\}\nonumber\\
 & &
  + \frac{\Lambda Bm}{6}
  \left[
   2(\cot\phi_R\csc\phi_R - \cot\phi_S\csc\phi_S)
   \right.\nonumber\\
  & & \left.+ \cos\phi_R (2 - \cos 2\phi_R)
   - \cos\phi_S (2 - \cos 2\phi_S)
   \right]
  + {\cal O}(m^3, \Lambda^2, \Lambda m^2).
  \label{eq:SdSangle6}
\end{eqnarray}
We mention here that although four angles $\varepsilon_S$, $\varepsilon_R$, 
$E_S$ and $E_R$ include the ${\cal O}(\Lambda)$ terms, the ${\cal O}(\Lambda)$ 
terms do not appear in Eq. (\ref{eq:SdSangle6}). Further, among the four 
angles $\varepsilon_S$, $\varepsilon_R$, $E_S$ and $E_R$, 
$\varepsilon_S$ and $\varepsilon_R$ can be considered as the 
observable angle on the curved space
\footnote{Angle $\varepsilon_R$ can be directly observed by 
the receiver $R$ (the observer of upcoming light ray).
Although $\varepsilon_S$ cannot be measured directly, if the position of
$S$ is determined precisely beforehand, $\varepsilon_S$ can also be
regarded as the observable angle.}.
However comparing Eq. (\ref{eq:SdSangle3}) to Eq. (\ref{eq:SdSangle6}), 
twice of $\varepsilon_S, \varepsilon_R$ or $\varepsilon_S - \varepsilon_R$ 
does not coincide simply with $\alpha_{\rm SdS}$ due to the presence of
$E_S$ and $E_R$.
\subsubsection{Integral Formula}
We now compute the total deflection angle $\alpha$ by using the integral formula,
Eq. (\ref{eq:angle9}). In the case of the Schwarzschild--de Sitter spacetime,
the Gaussian curvature $K^{\rm SdS}$ becomes
\begin{eqnarray}
 K^{\rm SdS} = - \frac{2m}{r^3}
  \left(1 - \frac{3m}{2r} + \frac{\Lambda}{6m}r^3
   - \Lambda r^2\right),
  \label{eq:KSdS}
\end{eqnarray}
and the areal element $d\sigma$ is
\begin{eqnarray}
  d\sigma = r
  \left(1 - \frac{2m}{r} - \frac{\Lambda}{3}r^2\right)^{-\frac{3}{2}}
  drd\phi,
  \label{eq:dSSdS}
\end{eqnarray}
In the same way that Eq. (\ref{eq:Schangle7}) was integrated,
we construct the quadrilateral $\Sigma^4$ bounded by three geodesics
$\gamma$, $\gamma_S$ and $\gamma_R$ and the unperturbed reference line 
$C_{\Gamma}$ which is the null geodesic $\Gamma$ on the de Sitter spacetime. 
We carry out the integration up to the order ${\cal O}(m^2, \Lambda, \Lambda m)$:
\begin{eqnarray}
 \iint_{\Sigma^4}K^{\rm SdS}d\sigma
  &=&
  - \int_{\phi_S}^{\phi_R}
  \int_{r_{\gamma}}^{r_{C_{\Gamma}}}
  \left(
   \frac{2m}{r^2} + \frac{3m^2}{r^3}
   + \frac{\Lambda}{3}r
  \right)drd\phi + {\cal O}(m^3, \Lambda^2, \Lambda m^2)
  \nonumber\\
 &=&
  - \frac{m^2}{8B^2}
  \left[
   24(\phi_R - \phi_S) + 4(\sin 2\phi_R - \sin 2\phi_S)
  \right]\nonumber\\
 & &
  + \frac{\Lambda Bm}{3}
  \left(\cot\phi_R\csc\phi_R - \cot\phi_S\csc\phi_S\right)
  + {\cal O}(m^3, \Lambda^2, \Lambda m^2),
  \label{eq:SdSangle7}
\end{eqnarray}
where $1/r_{C_{\Gamma}} = \sin\phi/B$ and $r_{\gamma}$ is given by
Eq. (\ref{eq:SdStrajectory1}); but in this case, it is enough to adopt 
the order ${\cal O}(m)$ solution of $r_{\gamma}$ as in Eq. (\ref{eq:Schangle7}).

The geodesic curvature $\kappa_g^{\rm SdS}$ is
\begin{eqnarray}
 \kappa_g^{\rm SdS} = \frac{b}{r^2}\left(1 - \frac{3m}{r}\right)
  \sqrt{1 - \frac{2m}{r} - \frac{\Lambda}{3}r^2} + \frac{d\Phi}{dt},
  \label{eq:kgSdS}
\end{eqnarray}
noting that the constant appearing in $\kappa_g^{\rm SdS}$ is $b$
instead of $B$. The line integral of $\kappa_g^{\rm SdS}$
is integrated along the curve $C_{\Gamma}$; then we have
\begin{eqnarray}
 \int_{C_{\Gamma}}\kappa_g^{\rm SdS}dt
  &=&
  \int_{\phi_S}^{\phi_R}
  \left(
   1 - \frac{2m}{r_{C_{\Gamma}}} - \frac{3m^2}{2r^2_{C_{\Gamma}}}
   + \frac{\Lambda}{6}r^2_{C_{\Gamma}}
	      \right)d\phi
  + \int_S^R\frac{d\Phi}{dt}dt
  \nonumber\\
  & &+ {\cal O}(m^3, \Lambda^2, \Lambda m^2)
  \nonumber\\
 &=&
  \phi_R - \phi_S + \frac{2m}{B}(\cos\phi_R - \cos\phi_S)
  \nonumber\\
  & &- \frac{m^2}{8B^2}
   \left[6(\phi_R - \phi_S) - 3(\sin 2\phi_R - \sin 2\phi_S)\right]
   \nonumber\\
 & &- \frac{\Lambda B^2}{6} (\cot\phi_R - \cot\phi_S)
  + \Phi_R - \Phi_S
  + {\cal O}(m^3, \Lambda^2, \Lambda m^2),
  \label{eq:SdSangle8}
\end{eqnarray}
where, the line element of the first term, $dt$, is replaced by $d\phi$ as follows:
\begin{eqnarray}
 dt = \frac{r^2}{b}
  \left(1 - \frac{2m}{r} - \frac{\Lambda}{3}r^2\right)^{-1}d\phi,
  \label{eq:dtSdS}
\end{eqnarray}
from Eq. (\ref{eq:dpdt}). We notice again that the constant in 
Eq. (\ref{eq:dtSdS}) is not $B$ but $b$.

Substituting Eqs. (\ref{eq:Schangle9}) and (\ref{eq:SdSangle5}) into 
Eq. (\ref{eq:SdSangle8}) yields
\begin{eqnarray}
  \int_{C_{\Gamma}} \kappa_g^{\rm SdS}dt
   &=&
   \frac{m}{B}
   \left[
    2(\cos\phi_R - \cos\phi_S)
    + \sin^2\phi_R\cos\phi_R - \sin^2\phi_S\cos\phi_S
   \right]\nonumber\\
 & &
  - \frac{m^2}{8B^2}
  \left\{
   6(\phi_R - \phi_S) - 3(\sin 2\phi_R - \sin 2\phi_S)
     \right.\nonumber\\
 & &
   \left. -4\left[
       \sin^3\phi_R\cos\phi_R(2 - \cos 2\phi_R)
       -
       \sin^3\phi_S\cos\phi_S(2 - \cos 2\phi_S)
      \right]
   \right\}
   \nonumber\\
 & &+ \frac{\Lambda Bm}{6}
  \left[
   \cos\phi_R (2 - \cos 2\phi_R)
   -
   \cos\phi_S (2 - \cos 2\phi_S)
	  \right]
  + {\cal O}(m^3, \Lambda^2, \Lambda m^2).
  \nonumber\\
  \label{eq:SdSangle9}
\end{eqnarray}
We note that the order ${\cal O}(\Lambda)$ terms cancel out 
in Eq. (\ref{eq:SdSangle9}). Then, from Eqs. (\ref{eq:SdSangle7}), 
(\ref{eq:SdSangle9}) and (\ref{eq:angle9}), the integral formula 
gives the same result as the angular formula:
\begin{eqnarray}
 \alpha_{\rm SdS} &=& - \iint_{\Sigma^4}K^{\rm SdS}d\sigma
  - \int_{C_{\Gamma}}\kappa_g^{\rm SdS}dt
  \nonumber\\
 &=&
  - \frac{m}{B}
  \left[
   2(\cos\phi_R - \cos\phi_S)
   + \sin^2\phi_R\cos\phi_R - \sin^2\phi_S\cos\phi_S
  \right]\nonumber\\
  & &+ \frac{m^2}{8B^2}
  \left\{
   30(\phi_R - \phi_S) + \sin 2\phi_R - \sin 2\phi_S
   \right.\nonumber\\
   & &\left.-4\left[
      \sin^3\phi_R\cos\phi_R (2 - \cos 2\phi_R)
      -
      \sin^3\phi_S\cos\phi_S (2 - \cos 2\phi_S)
     \right]
      \right\}\nonumber\\
 & &
  + \frac{\Lambda Bm}{6}
  \left[
   2(\cot\phi_R\csc\phi_R - \cot\phi_S\csc\phi_S)
   \right.
   \nonumber\\
   & &+ \left.\cos\phi_R (2 - \cos 2\phi_R)
   - \cos\phi_S (2 - \cos 2\phi_S)
  \right]
  + {\cal O}(m^3, \Lambda^2, \Lambda m^2).
   \label{eq:SdSangle10}
\end{eqnarray}
It should be pointed out here that Eqs (\ref{eq:SdSangle6}) and
(\ref{eq:SdSangle10}) do not contain the ${\cal O}(\Lambda)$ terms.
This result can be regarded as reflecting the fact the light ray 
does not bend in the de Sitter spacetime.

Here, we comment on the difference between the present results,
Eqs. (\ref{eq:SdSangle6}) and (\ref{eq:SdSangle10}), and 
our previous works, that is \cite{ak2012,arakida2016}.
In \cite{ak2012}, we showed that the equation of light trajectory 
does not change its form in the Schwarzschild and Schwarzschild--de
Sitter spacetime, and from the similarity of trajectory equation of light
ray, we concluded that the total deflection angle of light ray $\alpha$ also 
does not change even if $\Lambda \ne 0$, see Eq. (13) in \cite{ak2012}.
However, because the equation of light trajectory and its differential
equation give the relation of ``coordinate values $r$ and $\phi$'' and
generally they do not have the meaning of length and angle in the curved 
spacetime. Then the total deflection angle must be determined
by the metric in such a way that Eqs. (\ref{eq:angle4}) and (\ref{eq:angle5})
with Eq. (\ref{eq:fSdS}). Therefore, the total deflection angle $\alpha$ 
should contain the contribution of $\Lambda$.
On the other hand, in \cite{arakida2016} the contribution of $\Lambda$
appears as the ${\cal O}(\Lambda)$ term (see Eq. (25) in \cite{arakida2016}).
But the appearance of the ${\cal O}(\Lambda)$ term means that 
the light ray bends in the de Sitter spacetime and it is contrary to 
the conclusion in \ref{sec:desitter}. We can say that in \cite{arakida2016} 
the background is the Minkowski spacetime instead of the de Sitter one.
\subsection{Estimation of Observability}
Before closing this section, let us investigate the observability of
light deflection by the cosmological constant $\Lambda$ in terms of
the Schwarzschild--de Sitter solution.
From Eqs. (\ref{eq:SdSangle6}) and (\ref{eq:SdSangle10}),
we extract only the part of the order ${\cal O}(\Lambda m)$ terms and put
\begin{eqnarray}
 \alpha_{\rm SdS}^{\Lambda}
  &=&
  \frac{\Lambda Bm}{6}
  \left[
   2(\cot\phi_R\csc\phi_R - \cot\phi_S\csc\phi_S)
   \right.\nonumber\\
   & &\left.+ \cos\phi_R (2 - \cos 2\phi_R)
   - \cos\phi_S (2 - \cos 2\phi_S)\right].
  \label{eq:SdSangle11}
\end{eqnarray}
To evaluate $\alpha^{\Lambda}_{\rm SdS}$, we assume
$\Lambda \approx 10^{-52} ~ {\rm m}^{-2}$. As an example, let us consider
the galaxy as the lens object with mass $m \approx 10^{12}GM_{\odot}/c^2$
(where $G = 6.674 \times 10^{-11} ~{\rm m^3 \cdot kg^{-1} \cdot s^{-2}}$
is a Newtonian gravitational constant, $c = 3.0 \times 10^{8}~{\rm m}^2$ is
the speed of light in vacuum, and $M_{\odot} = 2.0 \times 10^{30} ~{\rm kg}$
is the mass of the sun) and we employ its radius as the impact parameter
$b \approx B \approx R_{\rm galaxy} \approx 5.0 \times 10^4 ~ {\rm ly}
\approx 5 \times 10^{20}~{\rm m}$. Using these values, we find
\begin{eqnarray}
 \frac{4m}{B} \approx 1.2 \times 10^{-5},\quad
  \frac{\Lambda Bm}{6} \approx 1.2 \times 10^{-17},
  \label{eq:SdSangle12}
\end{eqnarray}
and therefore in general $\alpha_{\rm SdS}^{\Lambda}$ is mostly 12 orders of 
magnitude smaller than the Schwarzschild part. However, due to the 
$\cot\phi\csc\phi$ term, $\alpha_{\rm SdS}^{\Lambda}$ increases sharply 
when the source $S$ and/or the receiver $R$ reaches the de Sitter horizon
$r \rightarrow r_{\Lambda} = \sqrt{3/\Lambda} \approx 1.73 \times 10^{26} ~{\rm m}$. 
We notice that in terms of the angular formula, the $\cot\phi\csc\phi$ 
terms come from the expression of $\varepsilon$ which is the observable angle
between the null geodesic $\gamma$ (actual path of light ray)
and the radial null geodesic ($\gamma_S$ or $\gamma_R$) on the 
optical reference geometry ${\cal M}^{\rm opt}$ or the curved 
$r, \phi$ subspace ${\cal M}^{\rm sub}$ (see Eq. (\ref{eq:SdSangle3})).
From $\sin\phi_{\Lambda} = B/r_{\Lambda}$, we have the angular coordinate 
of the de Sitter horizon as $\phi_{\Lambda} \approx 3.0 \times 10^{-6}$.
Now, we suppose the source $S$ and the receiver $R$ are placed in
a symmetrical position near the de Sitter horizon with respect
to the lens plane, and set $\phi_S = \phi_{\Lambda}$ and
$\phi_R = \pi - \phi_{\Lambda}$. 
Then $\alpha^{\Lambda}_{\rm SdS}$ becomes an extreme case:
\begin{eqnarray}
 \alpha^{\Lambda}_{\rm SdS} \approx - 5.5 \times 10^{-6}
  ~{\rm rad}.
  \label{eq:SdSangle13}
\end{eqnarray}
This value is almost half the value of the Schwarzschild part.
Perhaps the estimation, Eq. (\ref{eq:SdSangle13}), may largely exceed
the value restricted by the current cosmological observations, 
nevertheless from Eq. (\ref{eq:SdSangle11}),
if both the source $S$ and the receiver $R$ are located near the 
de Sitter horizon, we may be able to detect the contribution of 
the cosmological constant $\Lambda$ to light deflection, especially by
the gravitational lensing.
\begin{figure}[htbp]
\begin{center}
 \includegraphics[scale=0.6,clip]{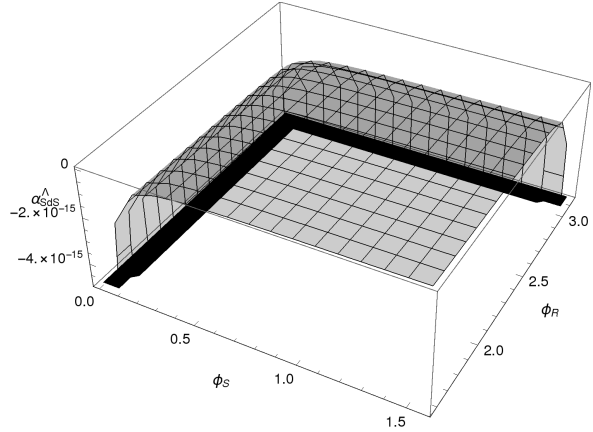}
 \caption{$\alpha^{\Lambda}_{\rm SdS}$
 as a function of $\phi_S$ and $\phi_R$. In this figure,
 $\alpha^{\Lambda}_{\rm SdS}$ is plotted for the domains
 $0 \leq \phi_S \leq \pi/2$ and $\pi/2 \leq \phi_R \leq \pi$.
 \label{fig:arakida-fig6}}
\end{center}
\end{figure}
Fig. \ref{fig:arakida-fig6} shows $\alpha^{\Lambda}_{\rm SdS}$
as a function of $\phi_S$ and $\phi_R$. It is found that
$\alpha^{\Lambda}_{\rm SdS}$ rapidly becomes a large value 
when $\phi_S \rightarrow 0$ and/or $\phi_R \rightarrow \pi$.
\section{Summary and Conclusions\label{sec:conclusion}}
In this paper, we re-examined the light deflection in the
Schwarzschild and the Schwarzschild--de Sitter spacetime. 
First, we proposed a definition of the total deflection angle $\alpha$
by constructing a quadrilateral $\Sigma^4$ on the optical reference 
geometry ${\cal M}^{\rm opt}$ determined by the optical
metric $\bar{g}_{ij}$ under the static and spherically 
symmetric spacetime. To construct a quadrilateral $\Sigma^4$, 
we introduced the unperturbed reference line $C_{\Gamma}$ 
which is the null geodesic $\Gamma$ on a background spacetime such 
as the Minkowski or the de Sitter spacetime, and $C_{\Gamma}$ is 
obtained by projecting $\Gamma$ vertically onto the curved $(r, \phi)$
subspace ${\cal M}^{\rm sub}$ After preparing the unperturbed 
reference line $C_{\Gamma}$, we laid a quadrilateral $\Sigma^4$ bounded 
by the null geodesic of the light ray $\gamma$, two radial null geodesics; 
$\gamma_S$ connecting the center $O$ and the source $S$ of 
the light ray and $\gamma_R$ connecting the center $O$ and the 
receiver $R$ of the light ray, and the unperturbed reference 
line $C_{\Gamma}$. The total deflection angle $\alpha$ is then 
defined as the sum of the four internal angles 
$\beta_p$ minus $2\pi$; see Eq. (\ref{eq:angle1}).

On the basis of the definition of the total deflection angle $\alpha$
and the Gauss--Bonnet theorem, we derived two formulas for calculating
the total deflection angle $\alpha$: (i) the angular formula
Eq. (\ref{eq:angle8}) that uses four angles determined on the optical
reference geometry ${\cal M}^{\rm opt}$ or the curved $(r, \phi)$
subspace ${\cal M}^{\rm sub}$ that is a slice of constant time $t$
and (ii) the integral formula, Eq. (\ref{eq:angle9}), on 
the optical reference geometry ${\cal M}^{\rm opt}$ is 
the areal integral of the Gaussian curvature $K$ in the area of 
the quadrilateral $\Sigma^4$ and the line integral of the geodesic
curvature $\kappa_g$ along the curve $C_{\Gamma}$. 
The angular formula, Eq. (\ref{eq:angle8}), is
derived directly from the definition of the total deflection angle, 
Eq. (\ref{eq:angle1}), which gives the meaning of the total deflection 
angle $\alpha$ geometrically clear. 

The angular formula can be used not only on the optical reference 
geometry ${\cal M}^{\rm opt}$ but also on the curved $(r, \phi)$ 
subspace ${\cal M}^{\rm sub}$ but an order ${\cal O}(m^n)$ solution 
for $r_{\gamma}$ is required when obtaining the total deflection 
angle $\alpha$ in order ${\cal O}(m^n)$. On the other hand, 
the integral formula should be calculated 
on the optical reference geometry ${\cal M}^{\rm opt}$; however, 
it is enough to prepare an order ${\cal O}(m^{n-1})$ solution for 
$r_{\gamma}$ when obtaining the total deflection angle $\alpha$ 
in order ${\cal O}(m^n)$.

We demonstrated that the two formulas give the same total deflection angle 
$\alpha$ for the Schwarzschild and the Schwarzschild--de Sitter spacetime. 
In particular, in the Schwarzschild case, the result coincides with 
Epstein--Shapiro's formula when the source $S$ and the receiver $R$ of 
the light ray are located at infinity, and in the Schwarzschild--de Sitter 
case, there appear order ${\cal O}(\Lambda m)$ terms as well as 
the Schwarzschild-like part, whereas order ${\cal O}(\Lambda)$ terms 
disappear because the light ray does not bend in the de Sitter spacetime.

In this paper, we took the stance that in order to obtain the total deflection
angle $\alpha$, we need to compare the two null geodesics $\gamma$ 
on the actual curved space and $\Gamma$ on the background space, and
for this purpose we vertically projected $\Gamma$ onto the curved space. 
Then, we expect that our approach is superior to that projecting 
$\gamma$ vertically onto the background space, e.g., the flat space 
for calculating the total deflection angle in the curved space.

\section{acknowledgments}
This work was supported by JSPS KAKENHI Grant Number 15K05089.

\appendix

\section{Relation of Coordinate Values among
Two Dimensional Curved spaces and three Dimensional Flat Plane
\label{append:space}}
The purpose of this appendix is to show the relation between coordinate 
values among the curved $(r, \phi)$ subspaces and the flat $(\rho, \varphi)$ 
plane under the static and spherical symmetric spacetime.

First, let us observe that two curved $(r, \phi)$ subspaces
${\cal M}^{\rm sub}_1$ and ${\cal M}^{\rm sub}_2$ characterized 
by Eq. (\ref{eq:metric3}) which is a slice of constant time $t$ of 
Eq. (\ref{eq:metric2}),
\begin{eqnarray*}
 d\ell^2 = \frac{1}{f(r)}dr^2 + r^2d\phi^2,
\end{eqnarray*}
have the same coordinate values. For the given curved $(r, \phi)$ subspaces
${\cal M}^{\rm sub}_1$ and ${\cal M}^{\rm sub}_2$, the form of $f(r)$
differs, i.e., $f_1(r_1) \ne f_2(r_2)$ But in the case of the static
and spherical symmetric spacetime, the radial coordinate $r$ is obtained 
as the circumference radius for $dr = 0$ (constant radius),
\begin{eqnarray}
 \ell = 2\pi r_1 = 2\pi r_2.
\end{eqnarray}
Then the curved $(r_1, \phi_1)$ subspace ${\cal M}^{\rm sub}_1$ 
and the curved $(r_2, \phi_2$ subspace ${\cal M}^{\rm sub}_2$ then 
take the same coordinate value:
\begin{eqnarray}
 r_1 = r_2,\quad  \phi_1 = \phi_2.
\end{eqnarray}
Next, in order to investigate the relation between the curved 
$(r, \phi)$ subspace ${\cal M}^{\rm sub}$ and the flat $(\rho, \varphi)$ 
plane, let us recall that the curved $(r, \phi)$ subspace
${\cal M}^{\rm sub}$ can be embedded into a three dimensional flat 
space with the cylindrical coordinates $(\rho, \varphi, z)$:
\begin{eqnarray}
 d\ell^2 = d\rho^2 + \rho^2d\varphi^2 + dz^2
  = \left[
     1 + \left(\frac{dz}{d\rho}\right)^2
    \right]d\rho^2 + \rho^2d\varphi^2.
  \label{eq:metric5}
\end{eqnarray}
As the same way above, the circumference $\ell$ can be determined as,
\begin{eqnarray}
 \ell = 2\pi r = 2\pi\rho,
\end{eqnarray}
then we find 
\begin{eqnarray}
 r = \rho,\quad \phi = \varphi.
\end{eqnarray}
This means that the flat $(\rho, \varphi)$ plane of the 
three-dimensional flat space and the curved $(r, \phi)$ subspace 
${\cal M}^{\rm sub}$ have the same coordinate values. 
We note that in the flat $(\rho, \varphi)$ plane, $\rho$ and $\varphi$ are 
directly related to the distance and angle, respectively, 
whereas in the curved $(r, \phi)$ subspace ${\cal M}^{\rm sub}$, 
$r$ and $\phi$ are just coordinate values; then, in general they do not 
simply mean the distance and angle.

From this consideration, we find that for the circumference $\ell$,
the flat $(\rho, \varphi)$ plane and the two curved $(r, \phi)$ subspaces, 
namely curved $(r, \phi)$ subspaces ${\cal M}^{\rm sub}_1$ and
${\cal M}^{\rm sub}_2$, have ``the same coordinate values'' 
$r_1 = r_2 = \rho$ and $\phi_1 = \phi_2 = \varphi$. This fact implies
the following; consider an arbitrary curve $C$, which may or 
may not be the geodesic. If the curve $C$ is given on one of the curved
$(r, \phi)$ subspaces or the flat $(\rho, \varphi)$ plane, its form 
does not change on another curved $(r, \phi)$ subspace by the
vertical projection, and vice versa.

For the purpose of reference, we obtain $z$ as a function of
$r$ or $\rho$. Noting that $r = \rho$, $z$ can be obtained by solving
the differential equation for the given $f(r)$:
\begin{eqnarray}
  1 + \left(\frac{dz}{dr}\right)^2 = \frac{1}{f(r)}.
\end{eqnarray}
For the Schwarzschild case, $f(r)$ is
\begin{eqnarray}
 f_{\rm Sch}(r) = 1 - \frac{2m}{r},
\end{eqnarray}
where $m$ is the central mass and $z$ is expressed as
\begin{eqnarray}
 z = 2\sqrt{2m(r - 2m)} = 2\sqrt{2m(\rho - 2m)}
\end{eqnarray}
which is well known as Flamm's paraboloid \cite{flamm1916}.
For the de Sitter case, the form of $f(r)$ is
\begin{eqnarray}
 f_{\rm dS}(r) = 1 - \frac{\Lambda}{3}r^2,
\end{eqnarray}
in which $\Lambda$ is the cosmological constant, and it yields
\begin{eqnarray}
 z = \sqrt{\frac{3}{\Lambda} - r^2} = \sqrt{\frac{3}{\Lambda} - \rho^2}.
\end{eqnarray}
For the Schwarzschild-de Sitter case, $f(r)$ becomes
\begin{eqnarray}
 f_{\rm SdS}(r) = 1 - \frac{2m}{r} - \frac{\Lambda}{3}r^2,
\end{eqnarray}
however, it is not easy to obtain the relation $z$
with $r$ or $\rho$ explicitly. Nonetheless it is obvious
that $z$ can be represented by a function of $r$ or $\rho$,
$z = z(r) = z(\rho)$.
\section{Note on Calculating Angle\label{append:angle}}
In this paper, we calculated the angles $\varepsilon_S$,
$\varepsilon_R$, $E_S$ and $E_R$ with the tangent formulas,
Eqs (\ref{eq:angle4}) and (\ref{eq:angle5}). However, it may seem to be
easier to perform the computation by using sine formula such as
Eq. (16) in \cite{ishihara_etal2016}:
\begin{eqnarray}
 \sin \Psi = \frac{b\sqrt{f(r)}}{r}
  \label{eq:sinchi}
\end{eqnarray}
which is equivalent to the cosine formula, see e.g.,
Eq. (15) in \cite{ishihara_etal2016}:
\begin{eqnarray}
 \cos \Psi = \frac{\sqrt{\gamma_{rr}}bA(r)}{r^2}\frac{dr}{d\phi} 
  = \frac{b}{r^2}\frac{dr}{d\phi},
  \label{eq:coschi}
\end{eqnarray}
where we replaced $A(r)$ by $f(r)$ and $\gamma_{rr}$ by $1/f^2(r)$ appeared
in \cite{ishihara_etal2016}. Because of the appearance of $f(r)$ in
Eq. (\ref{eq:sinchi}), the angle $\Psi$ is measured on the curved $(r, \phi)$ 
subspace ${\cal M}^{\rm sub}$ or the optical reference geometry 
${\cal M}^{\rm opt}$. In this appendix, we briefly show the formula,
Eq. (\ref{eq:sinchi}) diverges significantly when $\phi \rightarrow \pi/2$, 
whereas the tangent formulas, Eqs (\ref{eq:angle4}) and 
(\ref{eq:angle5}), avoid this divergence.

As a example, we calculate the angle $E$ of the Schwarzschild case
and for simplicity we obtain $E$ up to the order ${\cal O}(m)$.
From the sine formula, Eq. (\ref{eq:sinchi}), 
\begin{eqnarray}
 \sin E^{\rm sin} = \frac{b}{r_{C_{\Gamma}}}\sqrt{1 - \frac{2m}{r_{C_{\Gamma}}}}
 = \sin \phi - \frac{m}{b}\sin^2 \phi + {\cal O}(m^2),
\end{eqnarray}
where $1/r_{C_{\Gamma}} = \sin\phi/b$ then
\begin{eqnarray}
 E^{\rm sin} &=& \arcsin
  \left(
   \sin \phi - \frac{m}{b}\sin^2 \phi 
  \right)\label{eq:sinchi1}\\
 &=& \phi - \frac{m}{b}\sin \phi\tan\phi + {\cal O}(m^2).
  \label{eq:sinchi2}
\end{eqnarray}
Eq. (\ref{eq:sinchi2}) is equivalent to the expression of the second line
of Eq. (32) in \cite{ishihara_etal2016} after the following substitutions:
\begin{eqnarray}
 \arcsin(bu) = \phi,\quad
  - \frac{1}{2}br_g \frac{u^2}{\sqrt{1 - b^2u^2}}
   = - \frac{m}{b}\frac{b}{r}\frac{b}{\sqrt{r^2 - b^2}}
   = -\frac{m}{b}\sin\phi\tan\phi,
\end{eqnarray}
where $u = 1/r$ and $r_g = 2m$.

\begin{figure}[htbp]
\begin{center}
 \includegraphics[scale=0.7]{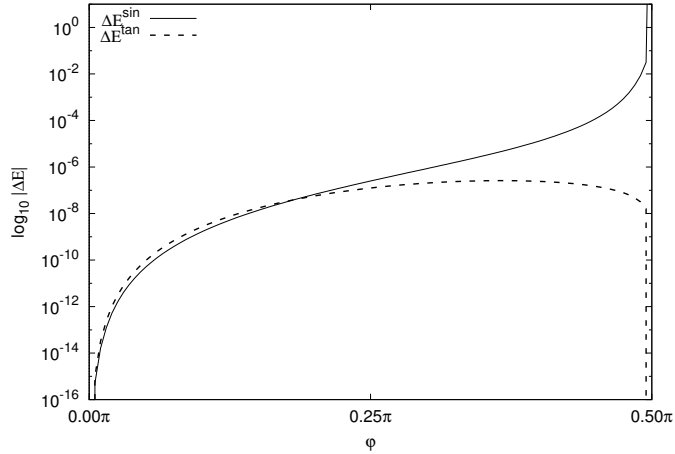}
 \caption{Deviations between Eqs. (\ref{eq:sinchi1}) and
 (\ref{eq:sinchi2}), $\Delta E^{\rm sin}$, (solid line), 
 and Eqs. (\ref{eq:tanchi1}) and (\ref{eq:tanchi2}), $\Delta E^{\rm tan}$,
 (dashed line). In this plot, we set $m/b = 0.001$. 
 \label{fig:arakida-fig7}}
\end{center}
\end{figure}

From the tangent formula, Eq. (\ref{eq:angle5}), $\tan E$ can be
expressed as
\begin{eqnarray}
 \tan E^{\rm tan} = \tan\phi - \frac{m}{b}\sin\phi\tan\phi
  + {\cal O}(m^2)
\end{eqnarray}
and
\begin{eqnarray}
 E^{\rm tan} &=& 
  \arctan\left(\tan\phi - \frac{m}{b}\sin\phi\tan\phi\right)
  \label{eq:tanchi1}\\
 &=& \phi - \frac{m}{b}\sin^2\phi\cos\phi + {\cal O}(m^2).
  \label{eq:tanchi2}
\end{eqnarray}
It is found that for $\phi \rightarrow 0$,
Eqs. (\ref{eq:sinchi2}) and (\ref{eq:tanchi2}) approach $0$,
whereas for $\phi \rightarrow \pi/2$, Eq. (\ref{eq:tanchi2}) reaches
$\pi/2$; on the other hand Eq. (\ref{eq:sinchi2}) becomes $-\infty$
due to the term $- (m/b)\sin\phi\tan\phi$.
Therefore, the angle should be computed with the tangent formula
rather than with the sine and cosine formulas.
Fig. \ref{fig:arakida-fig7} plots the absolute value of the deviation
between Eqs. (\ref{eq:sinchi1}) and (\ref{eq:sinchi2}), 
$\Delta E^{\rm sin}$, (solid line) and Eqs. (\ref{eq:tanchi1}) and 
(\ref{eq:tanchi2}), $\Delta E^{\rm tan}$, (dashed line).
It is clear that as the coordinate value $\phi$ becomes larger,
the result of Eq. (\ref{eq:sinchi2}) deviates from that of Eq. 
(\ref{eq:sinchi1}). On the other hand, the tangent formula 
does not produce such a divergence.


\begin{thebibliography}{99}
 \bibitem{dyson_etal1920} Dyson, F. W., Eddington, A. S., Davidson, C.:
	 A Determination of the Deflection of Light by the Sun's 
	 Gravitational Field, from Observations Made at 
	 the Total Eclipse of May 29, 1919,
	 Phil. Trans. Roy. Soc. of London. Series A,
	 {\bf 220}, 291--333 (1920)
 \bibitem{will2014} Will, C. M.: 
	 The Confrontation between General Relativity and Experiment,
	 Living Rev. Relativity, {\bf 17}, id. 4 (2014)
 \bibitem{schneider_etal1999} Schneider, P., Ehlers, J., Falco, E. E.:
	 Gravitational Lenses,
	 Springer Verlag, Berlin, Heidelberg, New York (1999)
 \bibitem{schneider_etal2006} Schneider, P., Kochanek, C., Wambsganss, J.: 
	 Gravitational Lensing: Strong, Weak and Micro,
	 Springer, Berlin, Heidelberg, New York (2006)
 \bibitem{islam1983} Islam, J. N.: 
	 The cosmological constant and classical tests of general relativity,
	 Phys. Lett. A, {\bf 97}, 239--241 (1983)
 \bibitem{rindler2007} Rindler, W., Ishak, M.:
	 Contribution of the cosmological constant to 
	 the relativistic bending of light revisited,
	 Phys. Rev. D, {\bf 76}, id. 043006 (2007)
 \bibitem{ishak2010} Ishak, M., Rindler, W.:
	 The relevance of the cosmological constant for lensing,
	 Gen. Rel. Grav., {\bf 42}, 2247--2268 (2010)
\bibitem{lake2002} Lake, K.: 
	Bending of light and the cosmological constant,
	Phys. Rev. D, {\bf 65}, id. 087301 (2002)
\bibitem{park2008} Park, M.: 
	Rigorous approach to gravitational lensing,
	Phys. Rev. D, {\bf 78}, id. 023014 (2008)
\bibitem{kp2008} Khriplovich, I. B., Pomeransky,A. A.:
	Does the Cosmological Term Influence Gravitational Lensing?,
	Int. J. Mod. Phys. D, {\bf 17}, 2255--2259 (2008)
\bibitem{sph2010} Simpson, F., Peacock, J. A., Heavens, A. F.:
	On lensing by a cosmological constant,
	On lensing by a cosmological constant, MNRAS, 
	{\bf 402}, 2009--2016 (2010)
\bibitem{bhadra2010} Bhadra, A., Biswas, S., Sarkar, K.:
	Gravitational deflection of light in the Schwarzschild-de Sitter 
	space-time,
	Phys. Rev. D, {\bf 82}, id. 063003 (2010)
\bibitem{miraghaei2010} Miraghaei, H., Nouri-Zonoz, M.:
	Classical tests of general relativity in the Newtonian 
	limit of the Schwarzschild-de Sitter spacetime,
	Gen. Rel. Grav. {\bf 42}, 2947--2956 (2010)
\bibitem{biressa2011} Biressa, T., de Freitas Pacheco, J. A.: 
	The cosmological constant and the gravitational light bending,
	Gen. Rel. Grav., {\bf 43}, 2649--2659 (2011)
 \bibitem{ak2012} Arakida, H., Kasai, M.: 
	 Effect of the cosmological constant on the bending of light 
	 and the cosmological lens equation,
	 Phys. Rev. D, {\bf 85}, id. 023006 (2012)
 \bibitem{hammad2013} Hammad, F.: 
	 a Note on the Effect of the Cosmological Constant on the Bending of Light,
	 Mod. Phys. Lett. A, {\bf 28}, id. 1350181 (2013)
\bibitem{lebedevlake2013} Lebedev, D., Lake, K.:
	On the influence of the cosmological constant on trajectories of 
	light and associated measurements in Schwarzschild de Sitter space,
	arXiv:1308.4931 (2013)
 \bibitem{batic_etal2015} Batic, D. Nelson, S., Nowakowski, M.:
	 Light on curved backgrounds,
	 Phys. Rev. D, {\bf 91}, id. 104015 (2015)
\bibitem{arakida2016} Arakida, H.: 
	Effect of the Cosmological Constant on Light Deflection: 
	Time Transfer Function Approach
	Universe, {\bf 2}, 5 (2016)
\bibitem{gibbons_werner2008a} Gibbons, G. W., Werner, M. C.:
	Applications of the Gauss Bonnet theorem to gravitational lensing,
	Class. Quant. Grav., {\bf 25}, id. 235009 (2008)
\bibitem{gibbons_werner2008b} Gibbons, G. W., Warnick, C. M., Werner, M. C.:
	Light bending in Schwarzschild de Sitter: 
	projective geometry of the optical metric,
	Class. Quant. Grav., {\bf 25}, id. 245009 (2008)
\bibitem{ishihara_etal2016} Ishihara, A., Suzuki, Y., Ono, T., Kitamura, T.,
	Asada, H.;
	Gravitational bending angle of light for finite distance and 
	the Gauss-Bonnet theorem,
	Phys. Rev. D, {\bf 94}, id. 084015 (2016)
\bibitem{ishihara_etal2017} Ishihara, A., Suzuki, Y., Ono, T. Asada, H.:
	Finite-distance corrections to the gravitational bending angle of 
	light in the strong deflection limit,
	Phys. Rev. D, {\bf 95}, id. 044017 (2017)
 \bibitem{weinberg1989} Weinberg, S.:
	 The cosmological constant problem,
	 Rev. Mod. Phys., {\bf 61}, 1--23 (1989)
 \bibitem{carroll2001} Carroll, S.:
	 The Cosmological Constant,
	 Living Rev. Relativity, {\bf 4}, id. 1 (2001)
 \bibitem{abramowicz1988} Abramowicz, M. A. Carter, B., Lasota, J. P.:
	 Optical reference geometry for stationary and static dynamics.,
	 Gen. Rel. Grav., {\bf 20}, 1173--1183 (1988)
 \bibitem{carroll2004} Carroll, S.:
	 Spacetime and Geometry: An Introduction to General Relativity,
	 Addison Wesley, San Francisco (2004)
\bibitem{klingenberg1978} Klingenberg, W.:
	A Course in Differential Geometry,
	Springer Verlag, New York (1978)
\bibitem{kreyszig1991} Kreyszig, E.:
	Differential Geometry,
	Dover Publications, New York (1991)
\bibitem{carmo2016} do Carmo, M. P.:
	Differential Geometry of Curves and Surfaces, 2nd ed.,
	Dover Publications, Mineola, New York (2016)
\bibitem{rindler2006} Rindler, W.:
	Relativity: Special, General, and Cosmological, 2nd ed., 
	Oxford University Press, New York (2006)
 \bibitem{epstein1980} Epstein, R., Shapiro, I. I.:
	 Post-post-Newtonian deflection of light by the Sun,
	 Phys. Rev. D, {\bf 22}, 2947--2949 (1980)
\bibitem{kottler1918} Kottler, F.:
	{\"U}ber die physikalischen Grundlagen der Einsteinschen 
	Gravitationstheorie,
	Annalen. Phys. {\bf 361}, 401--462 (1918)
\bibitem{flamm1916} Flamm, L.: 
	Beitr{\"a}ge zur Einsteinschen Gravitationstheorie,
	Phys. Z. {\bf 17}, 448--454 (1916)
\end{thebibliography}
\end{document}